%% file: main.tex
\documentclass[12pt]{article}
\usepackage{amsmath,amsfonts,amssymb}
\pdfoutput=1 
\usepackage{mathtools}
\usepackage{hyperref}
\usepackage{placeins}
\usepackage[bbgreekl]{mathbbol}
\usepackage{graphicx}
\usepackage{color}
\usepackage{epsfig}
\usepackage{psfrag}	
\usepackage{tikz}
\usetikzlibrary{snakes}
\usepackage{slashed}
\usepackage{ulem}
\usepackage[font=small,labelfont=bf,width=1\textwidth]{caption}
\usepackage{comment}
\usepackage[all]{xy}
\usepackage{multirow}
\usepackage{float}
\usepackage{tensor}
\usepackage {multirow}
\usepackage {booktabs}
\usepackage {fancyhdr}
\usepackage[T1]{fontenc}

\usepackage{tikz}
\usetikzlibrary{decorations.pathmorphing}
\usepackage{slashed}
\usepackage{ulem}
\usepackage{subfigure}
\usepackage{comment}
\usepackage{yhmath}
\usepackage[all]{xy}
\usepackage{multirow}
\usepackage{float}
\usepackage{mathtools}
\usepackage{xfrac}
\usepackage{wrapfig}
\usepackage{dsfont}
\usepackage{makecell}

\usepackage{color}

\newcommand \be{\begin{eqnarray}}
\newcommand \ee{\end{eqnarray}}

\numberwithin{equation}{section}
\DeclareMathOperator{\llangle}{\big\langle\hspace{-1.2mm}\big\langle\hspace{-.5mm}}
\DeclareMathOperator{\rrangle}{\hspace{-.5mm}\big\rangle\hspace{-1.2mm}\big\rangle}
\DeclareMathOperator{\Tr}{Tr}

\DeclareMathOperator{\sTr}{sTr}

\DeclareMathOperator{\sign}{sign}

\newcommand{\bea}{\begin{eqnarray}}
\newcommand{\eea}{\end{eqnarray}}
\newcommand{\beq}{\begin{equation}}
\newcommand{\eeq}{\end{equation}}
\newcommand{\bal}{\begin{equation}\begin{aligned}}
\newcommand{\eal}{\end{aligned} \end{equation}}

\newcommand{\cA}{{\mathcal A}}
\newcommand{\cB}{{\mathcal B}}

\newcommand{\cD}{{\mathcal D}}

\newcommand{\cL}{{\mathcal L}}

\newcommand{\cP}{{\mathcal P}}
\newcommand{\cQ}{{\mathcal Q}}

\newcommand{\cO}{{\mathcal O}}

\newmuskip\pFqmuskip

\newcommand{\address}[1]{\vbox{\center\em#1}}
\renewcommand{\title}[1]{\vbox{\center\huge{#1}}\vspace{5mm}}

\newcommand*\pFq[6][8]{%
  \begingroup % only local assignments
  \pFqmuskip=#1mu\relax
  \mathchardef\normalcomma=\mathcode`,
  \mathcode`\,=\string"8000
  \begingroup\lccode`\~=`\,
  \lowercase{\endgroup\let~}\pFqcomma
  {}_{#2}F_{#3}{\left[\genfrac..{0pt}{}{#4}{#5};#6\right]}%
  \endgroup
}
\newcommand{\pFqcomma}{{\normalcomma}\mskip\pFqmuskip}

\begin{document}
\begin{titlepage}
\begin{center}
\vspace*{20mm}

\title{Framed defects in ABJ(M)}
\vspace{5mm}
\renewcommand{\thefootnote}{$\alph{footnote}$}

Marco S. Bianchi$^a$, Luigi Castiglioni$^b$, Silvia Penati$^b$,\\
Marcia Tenser$^c$ and Diego Trancanelli$^d$%

\vskip 3mm

\address{
$^a$ Facultad de Ingeniería, Universidad San Sebastián,
Santiago, Chile
\\
$^b$ Dipartimento di Fisica, Universit\`a degli Studi di Milano--Bicocca and \\ INFN, Sezione di Milano--Bicocca, Piazza della Scienza 3, 20126 Milano, Italy
\\
$^c$ International Institute of Physics, Federal University of Rio Grande do Norte,\\ Campus Universit\'ario, Lagoa Nova, Natal-RN 59078-970, Brazil
\\
$^d$ Dipartimento di Scienze Fisiche, Informatiche e Matematiche,\\
Universit\`a di Modena e Reggio Emilia, via Campi 213/A, 41125 Modena, Italy and \\
INFN Sezione di Bologna, via Irnerio 46, 40126 Bologna, Italy}

\vskip 5mm

\tt{marco.bianchi@uss.cl, l.castiglioni8@campus.unimib.it, silvia.penati@mib.infn.it, marciatenser@gmail.com, diego.trancanelli@unimore.it}

\renewcommand{\thefootnote}{\arabic{footnote}}
\setcounter{footnote}{0}
\end{center}
\vspace{8mm}
\abstract{
\normalsize{
\noindent
We investigate the role of framing in a family of $1/24$ BPS Wilson loops in ABJ(M) theory, which define flows between $1/6$ BPS and the $1/2$ BPS superconformal fixed points. We analyze in perturbation theory how framing affects both the expectation values of these operators and the correlation functions of local insertions on the defect, as well as its interplay with RG flow and the $g$-theorem. We obtain a non-trivial identity between the one-point function of the defect stress tensor and a $\cQ$-exact correlator, which establishes a direct link between scale, superconformal invariance, cohomological equivalence and framing. This analysis confirms that framing one preserves supersymmetry and cohomological equivalence of operators at the quantum level, establishing a precise match with localization predictions.}
Finally, we propose a holographic interpretation of framing at strong coupling, identifying it with a coupling to the background $B$-field in the dual string theory.
}
\vfill
\end{titlepage}
\tableofcontents
%%%%%%%%%%%%%
\input{1intro}
\input{2theory}
\input{3Computation}
\input{4Corr}

\input{5Defect}
\input{6strongcoupling}
\input{Conclusions}
%%%%%%%%%%%%%
\bibliographystyle{utphys2}
\bibliography{refs}
\end{document}

%% file: 1intro.tex
\section{Introduction}

Supersymmetric Wilson loops define superconformal defect field theories, introducing one-dimensional extended objects that preserve part of the ambient supersymmetry and conformal invariance. As such, they offer a powerful framework to probe non-perturbative dynamics, symmetry breaking, and renormalization group flows in quantum field theories. 

In three-dimensional Chern-Simons-matter theories, particularly in ABJ(M) theory, supersymmetric Wilson loops are especially interesting due to their rich structure and deep connections to supersymmetric localization, integrability, and the AdS/CFT correspondence \cite{Drukker:2019bev}. They provide a fertile ground for exact computations and for exploring the interplay between geometry, topology, and quantum dynamics in gauge/string duality.

Among the various supersymmetric loop operators in ABJ(M) theory, a particularly rich family is provided by the $1/24$ BPS Wilson loops \cite{Castiglioni:2022yes}. These depend on a set of complex parameters which, despite supersymmetry, develop non-trivial beta-functions quantum mechanically. These trigger a supersymmetric RG flow which interpolates continuously between UV and IR superconformal fixed points, which are the known $1/6$ BPS and $1/2$ BPS configurations. Thus, this flow captures a network of defect RG flows and unifying several previously studied operators within a single framework.

The interpolating Wilson loops are cohomologically equivalent, differing by a ${\cal Q}$-exact deformation under a shared supercharge ${\cal Q}$. This equivalence suggests that their expectation values should coincide at the quantum level, provided supersymmetry is preserved throughout the regularization and renormalization procedure. However, a well-known subtlety arises in the context of Chern-Simons theory \cite{Witten:1988hf}: The non-invariance of the theory under a change of the trivialization of the normal bundle. This translates into the possibility of introducing a non-trivial framing phase which can be used to restore topological invariance at the quantum level and, in some schemes, to regularize short-distance singularities in the Wilson loop.
This procedure, which effectively displaces the integration contour to a nearby non-intersecting loop, introduces a framing dependence in the expectation value, manifesting as a topological phase proportional to the linking number between the original and displaced contours.

In the presence of matter fields, as in ABJ(M) theory, the theory is no longer topological and the framing dependence becomes more intricate. Unlike the one-loop exactness of framing phases in pure Chern-Simons theory, here higher-loop corrections can arise, and the resulting dependence may differ between the bosonic and fermionic sectors of the Wilson loop \cite{Bianchi:2016yzj}. Importantly, it has been argued \cite{Kapustin:2009kz} that only the choice of framing equal to one preserves supersymmetry at the quantum level and cancels the so-called cohomological anomaly \cite{Griguolo_2013a}. This implies that localization results, which rely on exact supersymmetry, compute expectation values precisely at this framing \cite{Kapustin:2009kz,Marino:2009jd,Drukker:2010nc}. 

The primary goal of this paper is to analyze the role of framing in interpolating Wilson loops and its consequences for defect RG flows in ABJ(M) theory. We highlight that the RG flow is independent of framing, as it origintes from short-distance local effects, whereas framing is a global and topological property of the loops. Building on the perturbative evaluation of framed diagrams of \cite{Bianchi:2024sod}, we perform a two-loop perturbative evaluation of the $1/24$ BPS Wilson loop at generic framing and for arbitrary values of the interpolating parameters. 

As our main result, we explicitly demonstrate that at framing one, the expectation value of the $1/24$ BPS operator becomes independent of the deformation parameters. While the cohomological equivalence of the interpolating family of operators was argued in \cite{Castiglioni:2022yes}, our calculation establishes that this equivalence indeed holds at the quantum level when the framing is set to one. Supersymmetric localization predicts the exact expectation value of the bosonic $1/6$ BPS Wilson loop, and cohomological equivalence then implies that all interpolating operators share this value. Our explicit two-loop computation confirms that framing one is precisely the point at which this chain of equivalences is realized.

When framing is different from one the Wilson loop expectation value of interpolating operators depends non-trivially on the coupling parameters, as well as on framing. 
Especially, for all operators but the 1/2 BPS case, framing does not merely appear as an overall phase which could be removed by considering their modulus. On the contrary, the two-node structure of the loops generically produces two terms with a relative phase difference. This produces interference and residual framing dependence when eliminating imaginary terms via a modulus. 

We further explore the influence of framing on physical observables localized on the defect. We study two-point correlation functions of operators on the Wilson loop and show that framing dependence persists away from the $1/2$ BPS point, where instead it cancels non-trivially. The special role of framing one also emerges computing  the one-point function of the defect stress tensor which we find to vanish at this specific value, where supersymmetry is restored. As a by-product, we obtain a Ward identity which links the expectation value of the defect stress tensor and a $Q$-exact correlator on the defect - see equations \eqref{eq:onept} and \eqref{eq:stresstensor1ptfunction}. These results establish a direct link between scale invariance, superconformal invariance, cohomological equivalence and framing one.

At framing one, the expectation values of the $1/24$ BPS operators coincide and become independent of deformation parameters, yet the defects still undergo non-trivial RG flow, as their beta functions remain nonzero. These beta functions are crucial in enforcing this parametric independence (see discussion between \eqref{eq:barevev} and \eqref{eq:renormalizedvev}). Since such expectation values define $g$-functions with monotonicity along RG flow, we examine the impact of framing on the $g$-theorem in defect CFTs. We find that the defect entropy can decrease, remain constant, or increase depending on whether the framing number is less than, equal to, or greater than one. Thus, at generic framing, these defects violate the $g$-theorem of \cite{Cuomo:2021rkm}, due to extra contributions in the Ward identities, originating from the framing-induced variation of the defect’s normal bundle under conformal transformations.

Finally, we propose a holographic interpretation of framing in the strong coupling regime, identifying it with the coupling of the dual string worldsheet to the background Kalb-Ramond $B$ field in the $\text{AdS}_4 \times \mathbb{CP}^3$ geometry. As in localization, this coupling naturally selects framing number equal to one, aligning with the expectation that the string solution probes supersymmetry preservation in a deeply quantum regime of the dual operators, requiring the absence of anomalies.

Altogether, our work elucidates the fundamental role of framing in the quantum dynamics of ABJ(M) Wilson loops, clarifies its interplay with supersymmetry and RG flows, and opens new avenues for understanding its effects in both field theory and string theory contexts. To summarize the main achievements of our work:
\begin{itemize}
    \item We compute the expectation value of $1/24$ BPS Wilson loops perturbatively at arbitrary framing.
    \item We establish that framing one is the appropriate supersymmetry preserving prescription for perturbative calculations of BPS loops.
    \item We explicitly show that at framing one cohomological equivalence holds at the quantum level, meaning that their expectation values are all captured by the same matrix model.
    \item Supersymmetry implies identities between defect correlation functions (e.g. vanishing of the stress tensor one-point function) which we find to hold at the quantum level precisely for $\mathfrak{f}=1$.
    \item We find that in general framing precludes a direct application of the $g$-theorem for defects in ABJ(M).
\end{itemize}

The rest of the paper is organized as follows. After section \ref{sec:theory} where we briefly review BPS Wilson loops and framing in ABJ(M) theory, in section \ref{sec:perturbative} we compute the most general 1/24 BPS Wilson loop at generic framing and for any value of the parameters. Up to two loops, we clarify the effects of framing on the loop integrals and discuss framing exponentiation. Examples of correlation functions on framed defects, and in particular the one-point function of the defect stress tensor, are computed in section \ref{sec:corrfunctions}. Section \ref{sec:gtheorem} is then devoted to the discussion of the $g$-theorem in the presence of framing. Finally, in section \ref{sec:strongcoupling} we discuss framing at strong coupling, exploiting the holographic description of Wilson loops in terms of minimal surface string configurations extending in ${\rm AdS}_2 \times {\mathbb{CP}}^3 \subset {\rm AdS}_4 \times {\mathbb{CP}}^3$. Framing is identified with a non-trivial Kalb-Ramond $B$-field, which the string may couple to. We provide a few arguments in favor of this identification. The paper closes with section \ref{sec:conclusions}, where we summarize the main results and highlight possible future developments. 

%% file: 2theory.tex
\section{Interpolating Wilson loops in ABJ(M) and framing}
\label{sec:theory}

We begin by briefly recalling the basics of BPS Wilson loops and framing in  Chern-Simons-matter theories. 

%%%%%%%%%%%%%%%%%%%

\subsection{BPS Wilson loops in ABJ(M)}
\label{subsec:abjmWLs}

Besides the usual gauge field holonomy, supersymmetric Wilson loops in Chern-Simons-matter theories may contain a coupling to the scalars of the theory only (the so-called {\it bosonic BPS loops}) or to both scalars and fermions (the so-called {\it fermionic BPS loops}).\footnote{A classification of the bosonic operators in ${\cal N}=4$ theories can be found in \cite{Drukker:2022bff}. The investigation of the fermionic ones started with \cite{Drukker:2009hy} for the ABJ(M) theory and was then generalized to less supersymmetric settings in \cite{Cooke:2015ila,Ouyang:2015qma,Ouyang:2015iza,Ouyang:2015bmy,Mauri:2017whf,Mauri:2018fsf}, see \cite{Drukker:2019bev} for a review.} The fermionic loops can be constructed via suitable deformations of the bosonic ones, as originally proposed in \cite{Mauri:2017whf, Drukker:2019bev} and further explored in \cite{drukker2020bps,Drukker:2020dvr,Drukker:2022ywj}. 

In ABJ(M) theory, which has ${\cal N}=6$ supersymmetry, the BPS Wilson loops may preserve different amounts of supercharges, from 1 (the 1/24 BPS loops) to 12 (the 1/2 BPS loops). In this paper we are especially interested in the 1/24 BPS loop introduced in \cite{Castiglioni:2022yes}. 
This operator depends on eight independent complex parameters $\alpha_i,\bar\alpha^i$ (with $i=1,2$) and $\bar\beta_j,\beta^j$ (with $j=3,4$). Bars do not stand for complex conjugation. The loop is supported on the circle 
\beq
\label{eq:circle0}
x^\mu = (\cos\tau,\sin\tau,0)
\, , \qquad \tau \in [0, 2\pi)
\eeq
and is defined as
\beq
\label{eq:W}
W_{1/24}=
\sTr \cP \exp\left(-i\oint \cL \,d\tau\right)\,,
\eeq
where $\mathcal{L}$ is the $U(N_1|N_2)$ superconnection\footnote{We indicate as $A_\mu, \hat{A}_\mu$ the gauge fields associated with the two nodes of the ABJ(M) quiver, $C_I, \bar{C}^I$ (with $I = 1, \dots , 4$) are the matter scalar fields in the (anti)fundamental representation of the $SU(4)$ R-symmetry group, whereas $\bar{\psi}^I$ and $\psi_I$ are their fermionic superpartners. }
\beq
\label{eq:1/24-superconnection}
\cL = \begin{pmatrix}
\cA+ \frac{1}{2} && \eta \, (\bar\alpha^1\bar{\psi}^2 - \bar\alpha^2 \bar{\psi}^1) + e^{-i\tau} \xi \, (\beta^3 \bar\psi^4 - \beta^4 \bar\psi^3) \\
\xi \, (\alpha_1 \psi_2 - \alpha_2 \psi_1) + e^{i\tau} \eta \, (\bar\beta_3 \psi_4 - \bar\beta_4 \psi_3) && \hat{\cA} 
\end{pmatrix}\,.
\eeq
The diagonal entries are given by
\beq
\label{eq:calA}
\cA = A_\mu \dot{x}^\mu - \frac{2 \pi i}{k}\vert\dot{x}\vert M_J^{\ I} C_I \bar{C}^J\,, \qquad
\hat\cA = \hat{A}_\mu \dot{x}^\mu - \frac{2 \pi i}{k}\vert\dot{x}\vert M_J^{\ I} \bar{C}^J C_I\, ,
\eeq
with scalar coupling matrix
\beq
\label{eqn:M124bos}
M_J^{\ I} = \begin{pmatrix}
-1 + 2\bar\alpha^1 \alpha_1 && 2\bar\alpha^1 \alpha_2 && 2e^{i\tau} \bar\alpha^1 \bar\beta_3 && 2e^{i\tau} \bar\alpha^1 \bar\beta_4 \\
2\bar\alpha^2\alpha_1 && -1 + 2\bar\alpha^2 \alpha_2 && 2e^{i\tau}\bar\alpha^2 \bar\beta_3 && 2e^{i\tau}\bar\alpha^2 \bar\beta_4 \\
2e^{-i\tau} \beta^3 \alpha_1 && 2e^{-i\tau} \beta^3 \alpha_2 && 1 + 2\beta^3 \bar\beta_3 && 2\beta^3 \bar\beta_4 \\
2e^{-i\tau} \beta^4 \alpha_1 && 2e^{-i\tau} \beta^4 \alpha_2 && 2\beta^4 \bar\beta_3 && 1+ 2\beta^4 \bar\beta_4
\end{pmatrix}\,.
\eeq
The off-diagonal entries in \eqref{eq:1/24-superconnection} contain fermionic couplings defined in terms of commuting spinors, $\eta$ and $\xi$, which on the circle \eqref{eq:circle0} read
\beq\label{eq:spinors}
\eta^\alpha = \sqrt{\frac{2\pi i}{k}}(1,-ie^{-i\tau})^\alpha \,,\qquad \xi^\alpha = \sqrt{\frac{2\pi i}{k}}(-ie^{i\tau},1)^\alpha \,.
\eeq

The definition in \eqref{eq:W}, in terms of the supertrace and the shift by $1/2$ in the first diagonal entry of ${\cal L}$ is equivalent \cite{Drukker:2019bev} to an alternative formulation with the trace and without the shift, which is more convenient in some instances, for example in perturbative computations.\footnote{How to perform perturbation theory in the formulation with the shift is explained in chapter 5 of \cite{Drukker:2019bev}.}

The operator $W_{1/24}$ interpolates among different supersymmetric representatives, obtained by setting the parameters in \eqref{eq:1/24-superconnection}-\eqref{eqn:M124bos} to specific values. By turning off either the $\{\alpha_i,\bar\alpha^i\}$ or the $\{\bar\beta_j,\beta^j\}$ parameters, the resulting operators become fermionic $1/6$ BPS Wilson loops \cite{Ouyang:2015iza,
Ouyang:2015bmy}. This specifies two possible branches of supersymmetric loops. For the first branch ($\bar\beta_j=\beta^j=0$), further setting $\alpha_i \bar\alpha^i =1$ enhances supersymmetry and the resulting operator is $1/2$ BPS. For the second branch ($\alpha_i=\bar\alpha^i=0$), the same happens when $\bar\beta_j \beta^j =-1$. We call $W_{1/2}$ the maximally supersymmetric Wilson loop. Finally, if we turn off all the parameters, we obtain the bosonic 1/6 BPS operator $W_{1/6}$ first introduced in \cite{Drukker:2008zx}. 

\begin{figure}[H]
    \centering
    \subfigure[]{\includegraphics[width=0.4\textwidth]{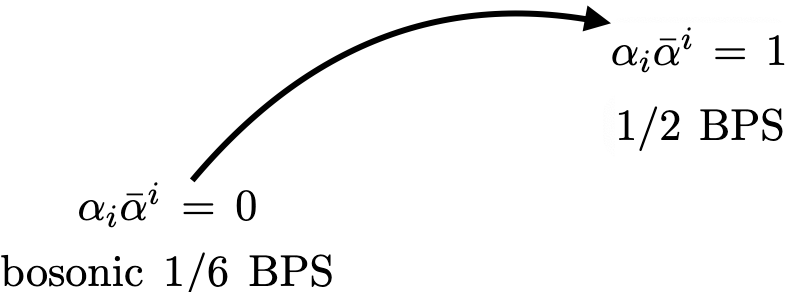}
    \label{subfig:flowbranch1}} 
    \qquad\subfigure[]{\includegraphics[width=0.4\textwidth]{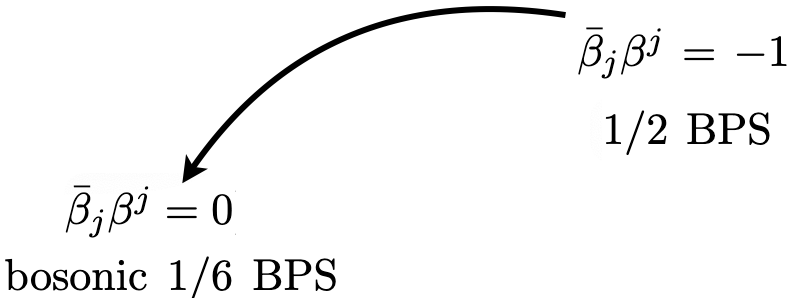}
    \label{subfig:flowbranch2}}  
    \caption{Schematic representation of the RG flows connecting fixed points. \subref{subfig:flowbranch1} In the first branch of solutions ($\bar\beta_j=\beta^j=0$), the 1/6 BPS bosonic loop is a UV unstable fixed point, while the 1/2 BPS fermionic loop is IR stable. \subref{subfig:flowbranch2} The opposite is true in the second branch of solutions ($\alpha_i=\bar\alpha^i=0$). Arrows are oriented from the UV to the IR.}
    \label{fig:flowscheme0}
\end{figure}

In \cite{Castiglioni:2022yes} it was found that the set of $\{\alpha_i,\bar\alpha^i,\bar\beta_j,\beta^j\}$ parameters undergoes a non-trivial renormalization, which in turn implies non-vanishing beta-functions. \footnote{See also \cite{Castiglioni:2023uus,Castiglioni:2023tci} and \cite{Castiglioni:2025iry} for a review.} At a difference with the 1/6 and 1/2 BPS cases, such UV divergences introduce a scale dependence which breaks explicitly the superconformal group classically preserved by these operators. At one loop in the ABJ(M) coupling constant, the beta-functions read 
\begin{equation}
\begin{split}
\label{eq:beta}
&\beta_{\alpha_i}=\frac{N_1+N_2}{2k}(\alpha_k \bar\alpha^k + \bar\beta_k \beta^k-1)\alpha_i \,, \quad \beta_{\bar\alpha^i}=\frac{N_1+N_2}{2k}(\alpha_k\bar\alpha^k + \bar\beta_k \beta^k -1)\bar\alpha^i\,, \\
&\beta_{\beta^j}=\frac{N_1+N_2}{2k}(\alpha_k \bar\alpha^k+ \bar\beta_k \beta^k+1)\beta^j \,, \quad \beta_{\bar\beta_j}=\frac{N_1+N_2}{2k}(\alpha_k\bar\alpha^k+ \bar\beta_k \beta^k+1)\bar\beta_j\, . 
\end{split}
\end{equation}
These beta-functions describe RG flows connecting different BPS loops. One can easily see \cite{Castiglioni:2022yes} that in the first branch the 1/6 BPS bosonic loop $W_{1/6}$ ($\alpha_i\bar\alpha^i=0$) sits at an unstable UV fixed point, whereas the $1/2$ BPS loop $W_{1/2}$ ($\alpha_i\bar\alpha^i =1$) sits at a stable IR fixed point. For the second branch, the opposite behavior is found: $W_{1/2}$ ($ \bar\beta_j \beta^j=-1$) is the UV fixed point, while $W_{1/6}$ ($ \bar\beta_j \beta^j=0$) is the IR fixed point. This is represented in figure \ref{fig:flowscheme0}, where the arrows indicate the direction of the flow, from the UV to the IR. In both cases, supersymmetry is partially preserved along the flows. In fact, RG trajectories connecting fixed points consist of representatives of $1/6$ BPS fermionic loops. For this reason, in \cite{Castiglioni:2022yes} they were dubbed {\it enriched flows}. 

BPS Wilson loops at the fixed points describe (un)stable superconformal defects, while Wilson loops at generic points along the RG flows correspond to supersymmetric but non-conformal defects obtained by perturbing the UV fixed point with marginally relevant operators. 

More general configurations exist, which correspond to RG flows driven by perturbations that break supersymmetry completely and connect BPS to non-BPS fixed points \cite{Castiglioni:2023uus}. Moreover, this construction can be generalized to interpolating circular Wilson loops defined on latitude contours on $S^2$ \cite{Castiglioni:2023tci}. 

The BPS Wilson loops we have mentioned here are cohomologically equivalent at the classical level. In fact, by construction, their definitions differ by a ${\cal Q}$-exact term, where ${\cal Q}$ is one of the mutually conserved supercharges. This implies that at the quantum level, in the absence of supersymmetry anomalies, their expectation values should coincide. 

%%%%%%%%%%%%%%%%%%%%%%%%%%%

\subsection{Framing in ABJ(M)}
\label{sec:framing}

In pure Chern-Simons (CS) theory, Wilson loops are famously connected to knot invariants \cite{Witten:1988hf}. 
Due to the topological nature of the theory, their expectation values are expected to be purely topological invariants. However, this requires introducing an additional phase counterterm called {\em framing}, which compensates for a topological anomaly induced by the regularization procedure of short-distance
singularities. 

Specifically, singularities may arise from the contraction of gauge fields evaluated on the same loop contour. A natural procedure to tame these singularities is point-splitting regularization. This introduces a framing: the choice of a non-intersecting nearby loop that slightly displaces the original contour \cite{Witten:1988hf,Guadagnini:1989am,Alvarez:1991sx}. 
Framing modifies the expectation value of a Wilson loop by a phase proportional to the linking number $\mathfrak{f}$ between the original contour and the regularized one, an integer given by the Gauss linking integral \cite{calugareanu1959integrale}
\begin{equation}\label{eq:Gauss}
   \mathfrak{f} =  \frac{1}{2\pi} \, \int_0^{2\pi} d\tau_1 \int_0^{\tau_1} d\tau_2 \frac{\dot x_1^\mu\dot x_2^{\nu}\epsilon_{\mu\nu\rho}x_{12}^\rho}{|x_{12}|^3}  \, .
\end{equation} 
As a result, the expectation value picks up an overall phase, which, for a $U(N)$ gauge group and a loop in the fundamental representation, reads
\begin{equation}\label{eq:CSframing}
    \langle W_{\text{CS}} \rangle_\mathfrak{f} = e^{\frac{i\pi N}{k}\mathfrak{f}}  \langle W_{\text{CS}} \rangle_{\mathfrak{f}=0}\,,
\end{equation}
where $k$ is the Chern-Simons level and $\langle W_{\text{CS}} \rangle_{\mathfrak{f}=0}$ indicates the framing independent part of the result.\footnote{This is nothing but the vacuum expectation value evaluated using ordinary dimensional regularization with dimensional reduction, which notoriously corresponds to zero framing.}
From a perturbative perspective, framing effects originate at one loop and exponentiate at higher orders, fully capturing their impact on the Wilson loop expectation value.

In the presence of matter, as in the ABJ(M) theory, framing effects arise in a more complicated way.\footnote{Framing also plays a central role in the perturbative computation of mesonic Wilson lines ending on fundamental matter, both bosonic and fermionic \cite{Gabai:2022vri,Gabai:2022mya}.} In particular, they are no longer one-loop effects.
In fact, for the bosonic 1/6 BPS Wilson loop it was found that the framing phase gets corrected at higher orders, as \cite{Bianchi:2016yzj}
\begin{equation}
\begin{split}
\label{eq:bosonic_framing}
\hspace{-0.2cm}
    \langle W_{1/6} \rangle_{\mathfrak{f}} \, & = \,  \frac{N_1}{N_1+N_2} e^{\frac{i\pi}{k}\left(N_1 - \frac{\pi^2}{2k^2}N_1N_2^2 + O(1/k^4) \right)\mathfrak{f}}\langle W_{\rm bos} \rangle_{\mathfrak{f}=0} \\
    &+ \frac{N_2}{N_1+N_2}e^{-\frac{i\pi}{k}\left(N_2 - \frac{\pi^2}{2k^2}N_1^2N_2 + O(1/k^4) \right)\mathfrak{f}}\langle \hat W_{\rm bos} \rangle_{\mathfrak{f}=0}
    \end{split}
\end{equation}
where $W_{\rm bos}$ and $\hat{W}_{\rm bos}$ are the single node bosonic loops whose connections are given by ${\cal A}$ and $\hat{\cal A}$ in \eqref{eq:calA}, respectively, with all the parameters set to zero. 

Supersymmetric localization predicts the exact non-perturbative expectation value of supersymmetric Wilson loops in ABJ(M) theory at $\mathfrak{f}=1$. The reason can be traced back to the fact that  this is the only supersymmetry preserving regularization scheme \cite{Kapustin:2009kz}. In fact, for $\mathfrak{f}=1$ the framing phase cancels exactly a cohomological anomaly \cite{Bianchi:2024sod}, supersymmetry is restored at the quantum level and cohomologically equivalent operators possess the same expectation value. In particular, this allows to evaluate the 1/2 BPS fermionic loop by computing the matrix model associated with the 1/6 BPS bosonic operator \cite{Marino:2009jd}. 

For the bosonic 1/6 BPS Wilson loop, the localization result expanded at weak 't Hooft couplings $N_1/k, N_2/k$ is in perfect agreement with the perturbative result \eqref{eq:bosonic_framing} evaluated at $\mathfrak{f}=1$. In the maximally supersymmetric case, {\it i.e.} the $1/2$ BPS Wilson loop, the localization result in the planar limit reads
\begin{equation}\label{eq:localizationresult}
    \langle W_{1/2} \rangle_{\mathfrak{f}=1} = \frac{1}{2}e^{\frac{i\pi}{k}(N_1-N_2)}\kappa\,,
\end{equation}
where $\kappa$ is a real function of $N_1/k, N_2/k$. 
A generalization of this result at generic framing can be attempted perturbatively, at weak coupling. This requires dealing with framing regularization not only for diagrams with bosonic propagators, but also for diagrams with fermion exchanges. The latter are more complicated to evaluate at non-trivial framing, due to the appearance of $(\eta, \xi)$ spinor bilinears, with $(\eta, \xi)$ defined in \eqref{eq:spinors}. Nevertheless, in \cite{Bianchi:2024sod} an efficient way to deal with spinor bilinears  has been introduced, which allows to isolate contributions potentially dependent on framing. Up to two loops, the Wilson loop expectation value at generic framing has been evaluated and reads \cite{Bianchi:2024sod}
\begin{equation}\label{eq:genericFrelation}
    \langle W_{1/2} \rangle_{\mathfrak{f}} = e^{\frac{i\pi}{k}(N_1-N_2)\mathfrak{f}}\langle W_{1/2} \rangle_{\mathfrak{f}=0}\,.
\end{equation}
The comparison between \eqref{eq:localizationresult} and \eqref{eq:genericFrelation} suggests that the phase $e^{\frac{i\pi}{k}(N_1-N_2)\mathfrak{f}}$ is plausibly the correct and exact generalization of  \eqref{eq:localizationresult} at generic framing number.\footnote{In principle, higher order corrections to the exponent proportional to $(\mathfrak{f}^2-1)$ could spoil the one-loop exactness of the phase. However, such corrections would imply that $\langle 
W_{1/2} \rangle_{\mathfrak{f}=0}$ is not real-valued.} A higher order calculation of $\langle W_{1/2} \rangle_{\mathfrak{f}}$ would be necessary to confirm the exactness of the framing phase at generic $\mathfrak{f}$, which is however beyond the scope of this paper. 

%% file: 3Computation.tex
\section{Perturbative evaluation at generic framing}
\label{sec:perturbative}

In this section, we compute the expectation value $\langle W_{1/24} \rangle_{\mathfrak{f}}$ of the 1/24 BPS interpolating loop \eqref{eq:W}-\eqref{eq:spinors} for generic framing and for generic values of the $\{\alpha_i,\bar\alpha^i,\bar\beta_j,\beta^j\}$ parameters. We carry out the computation up to two loops in perturbation theory using dimensional regularisation with dimensional reduction to tame short distance divergences. We stress that framing is not introduced here as a regulator of UV divergences, rather as a modification of the topology
of the loop. We use the techniques developed in \cite{Bianchi:2016yzj,
Bianchi_2016,
Bianchi:2018bke,Bianchi:2024sod} to evaluate diagrams with bosonic and fermionic propagators at non-trivial framing $\mathfrak{f}$. Setting $\bar\beta_j = \beta^j =0$ and $\alpha_i\bar\alpha^i=1$ or $\alpha_i\bar\alpha^i=0$, we expect to reproduce the results reviewed in the previous section for the 1/2 BPS fermionic loop and the 1/6 BPS bosonic one, respectively. 

The main interest in doing such an analysis lies in testing cohomological equivalence and matching with localization predictions at $\mathfrak{f}=1$. In fact, assuming that the matrix model yields expectation values at framing one and that, at this framing, cohomological equivalence is restored, irrespective of the value of the parameters, we should find that $\langle W_{1/24} \rangle_{\mathfrak{f}}$ at $\mathfrak{f}=1$ is independent of $\{\alpha_i,\bar\alpha^i,\bar\beta_j,\beta^j\}$  and matches the localization result for $\langle W_{1/6} \rangle_{\mathfrak{f}=1}$. In this section, we explicitly check this prediction perturbatively, finding complete agreement with localization at $\mathfrak{f}=1$ without resourcing to {\it ad hoc} introduction of framing phases.

%%%%%%%%%%%%%%%%%%%%%%%%%

\subsection{Perturbative results}

The two-loop evaluation of $\langle W_{1/24} \rangle_{\mathfrak{f}}$ at generic framing can be easily performed by combining the results in \cite{Bianchi:2024sod} and \cite{Castiglioni:2022yes}. Feynman diagrams are conveniently split into those devoid of matter field propagators, which are independent of the parameters, and those containing matter couplings, which carry a non-trivial dependence on $\{\alpha_i,\bar\alpha^i,\bar\beta_j,\beta^j\}$. 

The first set of diagrams provides the following four contributions, with wavy lines denoting gauge propagators, 
\begin{equation}\label{eq:gaugediagrams}
\begin{split}
    \vcenter{\hbox{\includegraphics[width=0.1\textwidth]{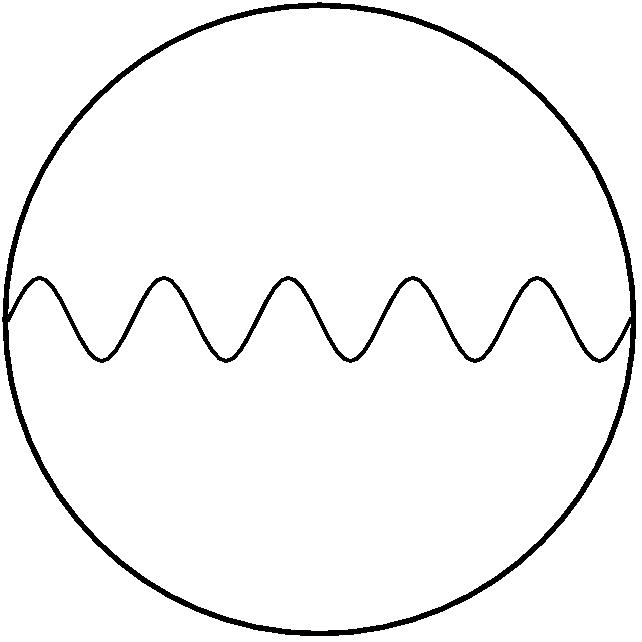}}} \quad &= \quad  \frac{i \pi}{k} \, (N_1^2-N_2^2) \, \mathfrak{f}\,, \hspace{3.15cm}
    \vcenter{\hbox{\includegraphics[width=0.1\textwidth]{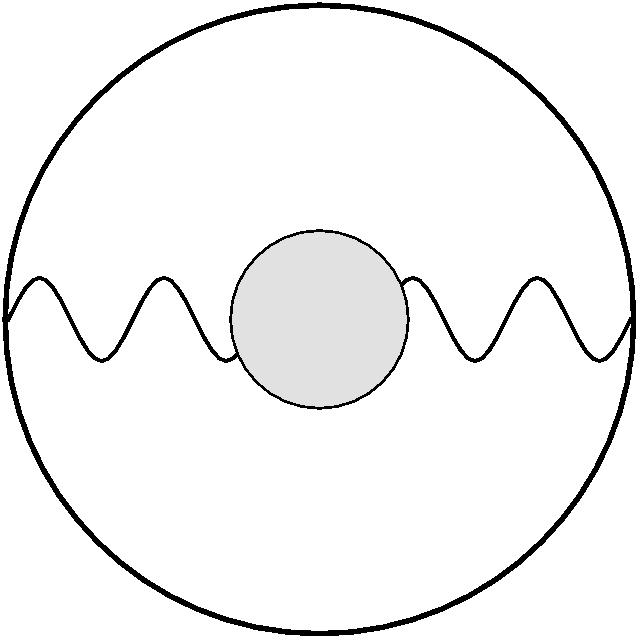}}}  \quad= \quad \frac{\pi^2}{k^2} \, N_1N_2(N_1+N_2) \,, \\
    \vcenter{\hbox{\includegraphics[width=0.1\textwidth]{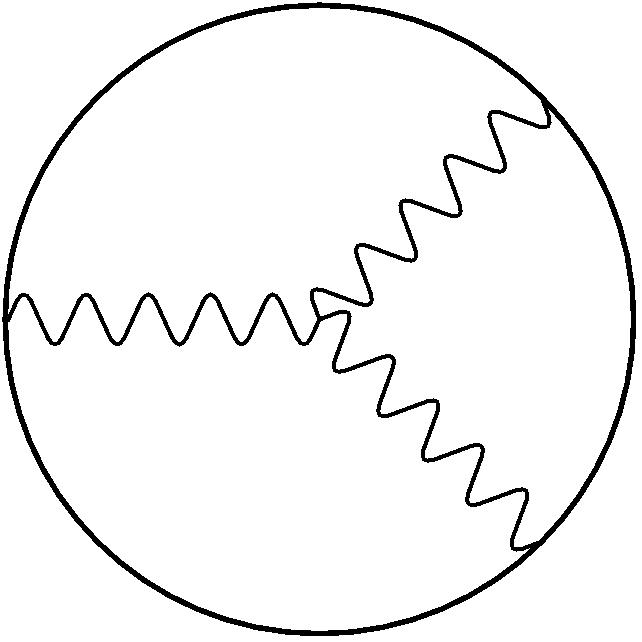}}}\quad &=\quad- \frac{\pi^2}{6k^2} \, \left(N_1^3-N_1+N_2^3-N_2\right)  \,, \qquad
    \vcenter{\hbox{\includegraphics[width=0.1\textwidth]{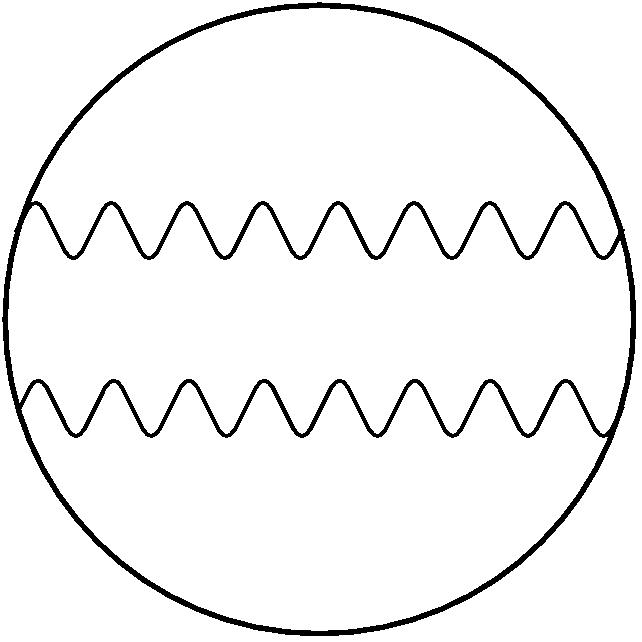}}} \quad= \quad - \frac{\pi^2}{2k^2} \, (N_1^3+N_2^3) \, \mathfrak{f}^2 \,.
\end{split}
\end{equation}
As is well known \cite{Guadagnini:1989am, Alvarez:1991sx}, framing dependence arises from diagrams containing collapsible propagators, {\it i.e.} free propagators whose end points can coincide. 

Diagrams in the second set include matter field propagators. They necessarily exhibit a non-trivial dependence on the parameters arising from the Wilson loop expansion in powers of the superconnection \eqref{eq:1/24-superconnection}. These diagrams can be easily evaluated by observing that after performing contractions the corresponding algebraic expression is given by an overall parameter-dependent factor times loop integrals at generic framing that are already known in the literature \cite{Bianchi:2024sod}.  
Therefore, we simply list the contribution from each non-vanishing diagram, where dashed (solid) lines denote scalar (fermionic) propagators.

\begin{equation}\label{eq:matterdiagrams}
\begin{split}
    \vcenter{\hbox{\includegraphics[width=0.1\textwidth]{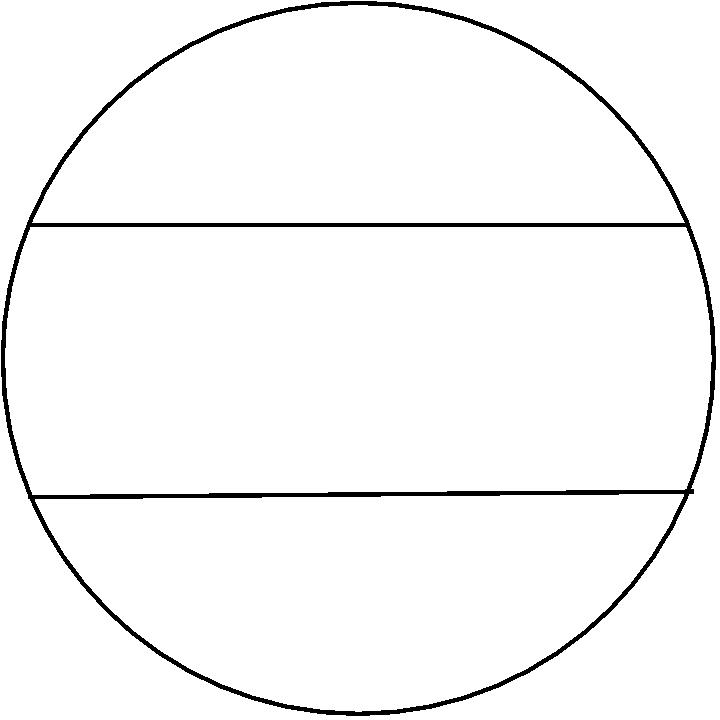}}}  \quad &=  \frac{\pi^2}{2k^2} \, N_1N_2(N_1+N_2) \, \left[ 3(\alpha_i\bar\alpha^i + \bar\beta_j \beta^j)^2 - (\alpha_i\bar\alpha^i - \bar\beta_j \beta^j)^2 \, \mathfrak{f}^2 \, \right] \,,\\
    \vcenter{\hbox{\includegraphics[width=0.1\textwidth]{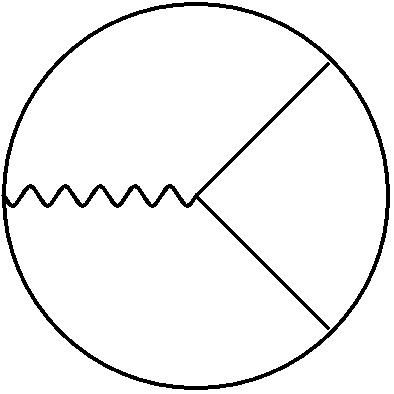}}}\quad &=- \frac{2\pi^2}{k^2} \, N_1N_2(N_1+N_2)(\alpha_i\bar\alpha^i - \bar\beta_j \beta^j)\,,\\
    \vcenter{\hbox{\includegraphics[width=0.1\textwidth]{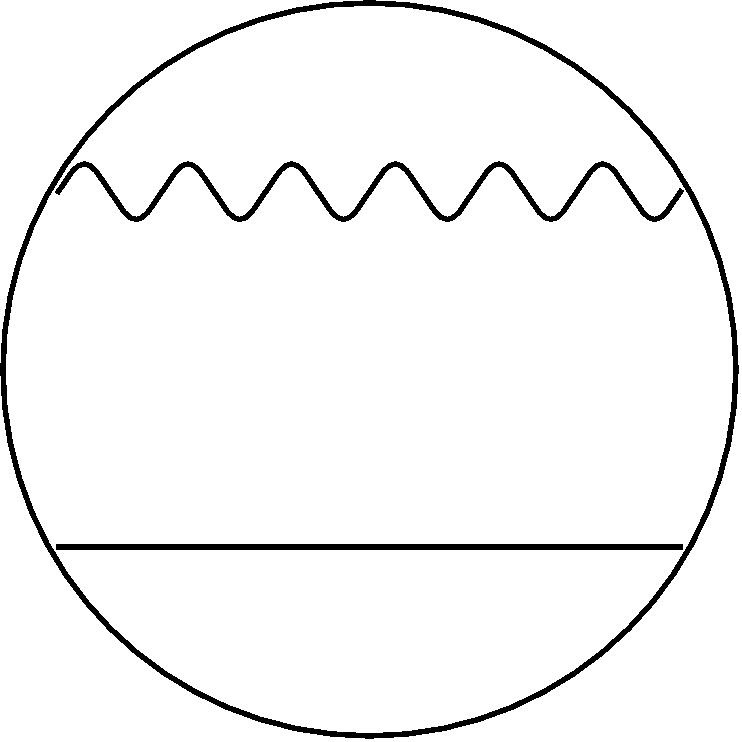}}} \quad &=  \frac{\pi^2}{k^2} \, N_1N_2(N_1+N_2) \,  (\alpha_i\bar\alpha^i - \bar\beta_j \beta^j) \, \mathfrak{f}^2  \,, \\
     \vcenter{\hbox{\includegraphics[width=0.1\textwidth]{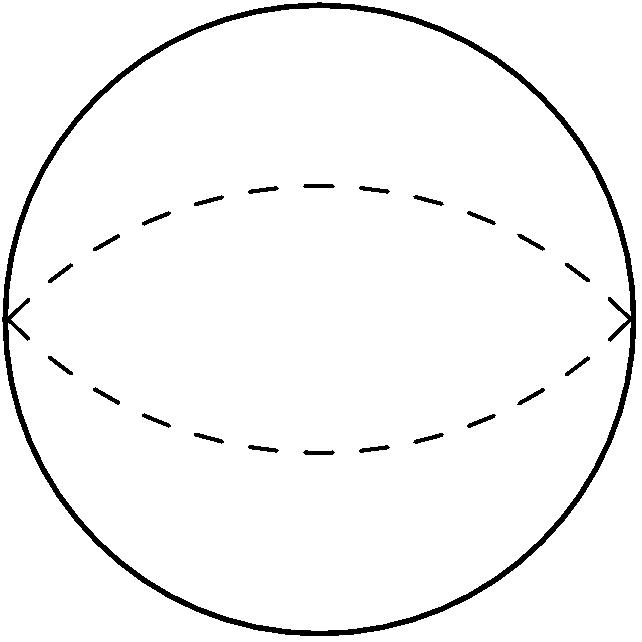}}} \quad&=  - \frac{2\pi^2}{k^2}N_1N_2(N_1+N_2)\alpha_i\bar\alpha^i \, \bar\beta_j \beta^j \,. 
\end{split}
\end{equation}
Once again, only diagrams with collapsible propagators contribute to framing at this order.

Combining the two partial results \eqref{eq:gaugediagrams} and \eqref{eq:matterdiagrams}, and normalizing the result by multiplying by $1/(N_1+N_2)$, we obtain the bare two-loop expectation value of the $1/24$ BPS loop at generic framing
\begin{align}
\label{eq:barevev}
    \langle W_{1/24} \rangle_\mathfrak{f} & = 1 + \frac{i\pi }{k}(N_1-N_2)\,\mathfrak{f}  + \frac{\pi^2}{6k^2}\bigg\{1- (N_1^2+N_2^2)(1+3\mathfrak{f}^2) - N_1N_2 \times \\ & \times \Big[ 
    12(\alpha\bar\alpha-2\alpha\bar\alpha \, \bar\beta\beta-\bar\beta\beta)-9(\alpha\bar\alpha-\bar\beta\beta)^2
     + 3\Big(\alpha\bar\alpha-\bar\beta\beta-2\Big)(\alpha\bar\alpha-\bar\beta\beta)\mathfrak{f}^2 -7 \Big]
 \bigg\}\,,\nonumber
\end{align}
where we have used the short-hand notation $\alpha\bar\alpha \equiv \alpha_i\bar\alpha^i$, $\bar\beta \beta \equiv \bar\beta_j \beta^j$.

This expression still contain bare parameters, thus it is not the final physical result. As already mentioned, the parameters undergo a non-trivial renormalization
\cite{Castiglioni:2022yes}, therefore in order to obtain the correct two-loop result they have to be replaced with their renormalized expressions. Precisely, 
\begin{equation}\label{eq:renormalization}
\alpha_i \rightarrow {\cal Z}_{\alpha} \, \alpha_i \, , \qquad 
\bar\alpha^i \rightarrow {\cal Z}_{\bar\alpha} \, \bar\alpha^i \, , \qquad
\beta^j \rightarrow {\cal Z}_{\beta} \, \beta^j \, , \qquad 
\bar\beta_j \rightarrow {\cal Z}_{\bar\beta} \, \bar\beta_j \, ,
\end{equation}
with the one-loop renormalization functions given by \cite{Castiglioni:2022yes} 
\begin{equation}
\label{eqn:renzeta}
\begin{split}
    &{\cal Z}_\alpha  = {\cal Z}_{\bar\alpha}  =  1+\frac{g^2}{8\pi\epsilon} \, (N_1+N_2)(\alpha\bar\alpha + \bar\beta\beta-1)  \,,  \\ & {\cal Z}_{\beta} = {\cal Z}_{\bar\beta} =  1+\frac{g^2}{8\pi\epsilon} \,(N_1+N_2) (\alpha\bar\alpha + \bar\beta\beta+1)  \, .
\end{split}
\end{equation}

As detailed in \cite{Castiglioni:2022yes}, at this order this has the effect of generating an additional contribution from the one-loop single fermion exchange diagram, which {\em per se} is order $\epsilon$ in dimensional regularization but it contributes at two loops when multiplied by the $1/\epsilon$ divergence encoded in the  renormalization functions of the parameters. Incorporating this additional term in the result \eqref{eq:barevev}, we finally obtain the two-loop renormalized expectation value at generic framing $\mathfrak{f}$
\begin{align}
\label{eq:renormalizedvev}
    \langle W_{1/24} \rangle_\mathfrak{f} =& 1 +\frac{i\pi }{k}(N_1-N_2)\,\mathfrak{f} \\& + \frac{\pi^2}{6k^2}\bigg[ 
    1-(N_1-N_2)^2 (1+3\mathfrak{f}^2) + 3N_1N_2 \, (\alpha\bar\alpha-\bar\beta\beta-1)^2(1-\mathfrak{f}^2) + 2N_1N_2 \bigg]\,.\nonumber 
\end{align}

First of all, we observe that for the particular choice $\alpha \bar\alpha = 1, \bar\beta\beta = 0$ (or $\alpha \bar\alpha = 0, \bar\beta\beta = -1$) this expression  reproduces exactly the result for the 1/2 BPS Wilson loop at generic framing, obtained in \cite{Bianchi:2024sod}. Furthermore, setting $\mathfrak{f}=1$ the result coincides with the localization prediction at weak coupling obtained by assuming exact cohomological equivalence with the 1/6 BPS bosonic operator \cite{Marino:2009jd}. As a further check, setting $\mathfrak{f}=0$ we reproduce the perturbative result of \cite{Bianchi:2013zda,Bianchi:2013rma,Griguolo_2013a}. 

Setting $\mathfrak{f}=1$ in \eqref{eq:renormalizedvev}, the result loses completely its dependence on the parameters. In other words, we can write
\begin{equation}
\label{eq:quantumcoho}
 \langle W_{1/24} (\alpha_i, \bar\alpha^i, \bar\beta_j, \beta^j) \rangle_{\mathfrak{f} =1} = \langle W_{1/24} (0,0,0,0) \rangle_{\mathfrak{f} =1} \equiv \langle W_{1/6} \rangle_{\mathfrak{f} =1}\, , \hskip .5cm \forall \;  \alpha_i, \bar\alpha^i, \bar\beta_j, \beta^j , 
\end{equation}
which is the quantum version of the cohomological equivalence. As expected, this is a non-trivial check that at framing one the cohomological anomaly gets canceled and supersymmetry is restored at the quantum level. The expectation value at framing one for any 1/24 BPS Wilson loop can then be computed using the matrix model associated with the bosonic 1/6 BPS operator. While this was already established for the $1/2$ BPS loop \cite{Bianchi:2024sod}, equation \eqref{eq:quantumcoho} extends the result to the entire interpolating family of $1/24$ BPS operators.

We stress that the renormalization of the parameters is crucial in drawing these conclusions. In fact, the unrenormalized expression \eqref{eq:barevev} does not lead to \eqref{eq:quantumcoho}.

Since beta-functions do not depend on framing, we can state that the RG flow pattern is basically the same for any value of $\mathfrak{f}$ and reproduces what has been found in \cite{Castiglioni:2022yes} for the zero framing case. On the other hand, an explicit dependence appears in the expectation value \eqref{eq:renormalizedvev}, which induces a change of $\langle W_{1/24} \rangle$ as a function of the renormalized parameters. We are going to discuss this point carefully in section \ref{sec:gtheorem}. 

%%%%%%%%%%%%%%%%%%%%%%%%%

\subsection{Framing exponentiation}

As reviewed in section \ref{sec:framing}, for the 1/2 BPS loops the matrix model predicts that framing contributions exponentiate to an overall one-loop exact phase. A more intricate structure appears for the 1/6 BPS bosonic loop, see \eqref{eq:bosonic_framing}, where the two blocks $W_{\rm bos}$ and $\hat{W}_{\rm bos}$ acquire different phases. Furthermore, in this case the two phases are corrected at higher orders \cite{Bianchi:2016yzj}. It is then interesting to investigate what happens along the RG flows, away from the two fixed points.

Our general result in \eqref{eq:renormalizedvev} seems to indicate that there is still a framing phase that might exponentiate. Whether it exponentiates to a single phase factor or it splits into different phase factors can be better understood by analyzing how framing contributions organize themselves in the calculation done using the one-dimensional auxiliary field method \cite{Dorn:1986dt,Craigie:1980qs,Castiglioni:2022yes}, where the expectation value of a BPS Wilson loop described by a superconnection $\cL$ is given by
\begin{equation}\label{eq:1dW}
    \langle W \rangle = \langle \Tr \Psi(2\pi) \bar\Psi(0)\rangle \qquad {\rm with } \quad \Psi = \begin{pmatrix}
        z & \varphi \\ \tilde\varphi & \tilde z
    \end{pmatrix}\, .
\end{equation}
Here $\Psi$ is a Grassmann-odd supermatrix consisting of one-dimensional fields which minimally couple to $\cL$. Performing the trace in \eqref{eq:1dW} and identifying equivalent contributions, we obtain that the expectation value can be calculated as the sum of the two-point functions~\cite{Castiglioni:2022yes} 
\begin{equation}\label{eq:1dW2}
    \langle W \rangle = \langle z(2\pi) \bar z(0) \rangle + \langle \tilde z(2\pi) \bar{\tilde z}(0) \rangle\,.
\end{equation}
In the case of the bosonic 1/6 BPS operator, this structure trivially reflects the block-diagonal nature of the connection, under the identification 
\begin{equation}
    \langle W_{\rm bos} \rangle = \langle z(2\pi) \bar z(0) \rangle\,, \qquad \langle \hat W_{\rm bos} \rangle=\langle \tilde z(2\pi) \bar{\tilde z}(0) \rangle\,.
\end{equation}
Therefore, according to the splitting in \eqref{eq:1dW2}, the appearance of two different framing phases in \eqref{eq:bosonic_framing} can be traced back to a different framing contribution to the $z$ and $\tilde{z}$ sectors. 

In the more general case of interpolating Wilson loops, even if the superconnection is no longer block-diagonal, the splitting \eqref{eq:1dW2} is still valid. Therefore, we expect that also in this case framing contributions will eventually sum up into two independent phases. From our two-loop result \eqref{eq:renormalizedvev}, we conjecture that 
\begin{equation}\label{eq:exponentialization}
\begin{split}
    \langle W_{1/24} \rangle_{\mathfrak{f}} &= \frac{N_1}{N_1+N_2} e^{\frac{i\pi}{k}[(N_1-(\alpha\bar\alpha-\bar\beta\beta)N_2) + O(1/k^2)] \, \mathfrak{f}} \, \langle z(2\pi) \bar z(0)\rangle_{\mathfrak{f}=0} \\ &+ \frac{N_2}{N_1+N_2} e^{\frac{i\pi}{k}[((\alpha\bar\alpha-\bar\beta\beta)N_1-N_2)+ O(1/k^2)]\, \mathfrak{f}} \, \langle \tilde z(2\pi) \tilde{\bar z}(0)\rangle_{\mathfrak{f}=0} \, ,
\end{split}
\end{equation}
where $\langle z(2\pi) \bar z(0)\rangle_{\mathfrak{f}=0}$ and 
$\langle \tilde z(2\pi) \tilde {\bar z}(0)\rangle_{\mathfrak{f}=0}$ are the framing zero contributions to the Wilson loop from the upper and lower diagonal blocks of \eqref{eq:1/24-superconnection}, respectively. For $\mathfrak{f}=1$, this expression is consistent with the exponentiation proposed in \cite{Castiglioni:2022yes}. Moreover, at the bosonic fixed point ($\alpha\bar\alpha=\bar\beta\beta=0$) we recover \eqref{eq:bosonic_framing}, whereas at the $1/2$ BPS fixed point ($\alpha\bar\alpha=1,\bar\beta\beta=0$ or $\alpha\bar\alpha=0,\bar\beta\beta=-1$) we reconstruct a single phase, as expected.

While the structure in \eqref{eq:exponentialization} can be trivially verified at one-loop order, we expect the exponentiation of framing to persist at higher loops. The reason is that each block independently sums up all diagrams containing collapsible  bosonic and fermionic propagators, which are sensitive to framing. Although higher-order corrections are expected to modify the phases in \eqref{eq:exponentialization}  — similarly to the $1/6$ BPS case — their split structure, which underlies the exponentiation in the form \eqref{eq:exponentialization}, should remain intact. 

In conclusion, for two-node WLs the framing dependence cannot be simply removed from the expectation values by taking the modulus, except for the special case of the $1/2$ BPS point.

%% file: 4Corr.tex
\section{Correlation functions in framed defects}\label{sec:corrfunctions}

The general procedure for computing (fermionic) Feynman diagrams at generic framing developed in \cite{Bianchi:2024sod} can be applied to perform a perturbative analysis of other physical quantities, in addition to Wilson loop expectation values. Here, we focus on the evaluation of correlation functions on framed Wilson loops. Our main interest is to understand how framing contributions affect 
correlation functions on a framed defect. As we discuss later, this has important implications for the defect theory, and may shed some light on the holographic interpretation of framing at strong coupling. Moreover, correlation functions of the defect stress tensor provide the most direct understanding of the connection between supersymmetry anomalies and framing. 

We recall that a $n$-point correlation function on a circular Wilson loop is defined as\footnote{We use the notation $x_i^\mu \equiv x^\mu(\tau_i)$. Moreover, $W(\tau_i,\tau_j)$ indicates an arcwise Wilson link parametrized by $\tau \in (\tau_i,\tau_j)$.}
\begin{equation}\label{eq:correlators}
    \llangle \cO_1(x_1)\cO_2(x_2)\ldots \cO_n(x_n)\rrangle \equiv  \frac{\langle \Tr W(2\pi, \tau_1) \cO_1(x_1) W(\tau_1,\tau_2)\cO_2(x_2)\ldots \cO_n(x_n)W(\tau_n,0) \rangle}{\langle W(2\pi,0) \rangle}\,.
\end{equation}

Focusing for simplicity on a two-point function, we look for  contributions that are framing-dependent. These are expected to come from diagrams that contain free collapsible propagators. Therefore, in doing a loop calculation it is convenient to split the diagrams into the collapsible and non-collapsible subsets and focus on the first class. 

In what follows, we first consider the simpler case of the bosonic 1/6 BPS Wilson loop describing a superconformal defect at the fixed point of the enriched RG flow, then we generalize to fermionic 1/6 BPS Wilson loops along the RG flows, including the 1/2 BPS fixed point. 

%%%%%%%%%%%%%%%%%%%%%%%%%%%%%

\subsection{Bosonic defect}

We start from the simpler case of the $W_{1/6}$ defect, and evaluate the two-point correlation function of a local diagonal supermatrix operator $\cal O$
\begin{equation}
    \llangle \cO(x_1)\cO(x_2) \rrangle_{1/6} = \frac{\langle \Tr W_{1/6}(2\pi,\tau_1)\cO(x_1)W_{1/6}(\tau_1,\tau_2)\cO(x_2)W_{1/6}(\tau_2,0) \rangle}{\langle W_{1/6}\rangle} \, .
\end{equation}
%To make an explicit computation, we choose $\cal O$ to be the scalar operator with only the gauge fields
We choose
\begin{equation}\label{eq:OperatorA}
    \cO(x) = \begin{pmatrix}
        \cB(x) & 0 \\ 0 & \hat\cB(x)
    \end{pmatrix} \, ,
\end{equation}
where $\cB, \hat\cB$ are  gauge covariant scalar operators localized on the defect.
%\begin{equation}\label{eq:OperatorA}
%    \cO(x) = \begin{pmatrix}
%        A_{\mu}\dot x^{\mu} & 0 \\ 0 & \hat A_{\mu}\dot x^{\mu}
%    \end{pmatrix} \, .
%\end{equation}
%whose tree level contribution to the two-point function is simply given by
%\begin{equation}
%    \llangle \cO(x_1)\cO(x_2) \rrangle_{1/6}^{(\rm tree)} = \dot{x}_1^{\mu}\dot{x}_2^{\nu} \, \frac{\langle A_{\mu}(x_1)A_{\nu}(x_2) \rangle + \langle \hat A_{\mu}(x_1) \hat A_{\nu}(x_2) \rangle}{\langle W_{1/6} \rangle}.
%\end{equation}

At one loop, diagrams potentially contributing to framing are planar, collapsible gauge corrections with both endpoints on the loop and no operator insertions separating them. Specifically, if such points as labelled by parameters $\tau_3$ and $\tau_4$,  
%shown in figure \ref{fig:disconnected}. 
%\begin{figure}[ht]
%    \centering
%    \subfigure[]{\includegraphics[width=0.25\textwidth]{img/diagram1.jpg}} 
%    \qquad\subfigure[]{\includegraphics[width=0.25\textwidth]{img/diagram2.jpg}}
%    \qquad\subfigure[]{\includegraphics[width=0.25\textwidth]{img/diagram3.jpg}}
%    \caption{One-loop diagrams that contain collapsible propagators.}
%    \label{fig:disconnected}
%\end{figure}
all framing dependent diagrams evaluate to the same integrand, but display different integration regions
\begin{equation}
    \left( \int_0^{\tau_1}d\tau_3 \int_0^{\tau_3} d\tau_4 + \int_{\tau_1}^{\tau_2} d\tau_3 \int_{\tau_1}^s{\tau_3} d\tau_4 + \int_{\tau_2}^{2\pi} d\tau_3 \int_{\tau_2}^{\tau_3} d\tau_4  \right) \frac{\epsilon_{\mu\nu\rho}\dot x_3^{\mu} \dot x_4^{\nu} (x_3-x_4)^{\rho}}{|x_3-x_4|^3}\,.
\end{equation}
Combining them and adding suitable framing independent and vanishing contributions
\begin{equation}
    \left( \int_{\tau_2}^{2\pi}d\tau_3 \int_{0}^{\tau_1} d\tau_4 + \int_{\tau_2}^{2\pi} d\tau_3 \int_{\tau_1}^{\tau_2} d\tau_4 + \int_{\tau_1}^{\tau_2} d\tau_3 \int_{0}^{\tau_1} d\tau_4  \right) \frac{\epsilon_{\mu\nu\rho}\dot x_3^{\mu} \dot x_4^{\nu} (x_3-x_4)^{\rho}}{|x_3-x_4|^3} = 0\,,
\end{equation}
we manage to factorize the operator insertions and the framing dependent corrections, reconstructing the full Gauss linking integral \eqref{eq:Gauss}. Schematically, we obtain for the upper block a one-loop framing dependent correction of the form
\begin{equation}
    \sim\langle \cB(x_1)  \cB(x_2) \rangle^{(0)}\times \int_{0}^{2\pi} d\tau_3 \int_{0}^{\tau_3} d\tau_4 \frac{\epsilon_{\mu\nu\rho}\dot x_3^{\mu} \dot x_4^{\nu} (x_3-x_4)^{\rho}}{|x_3-x_4|^3} \,.
\end{equation}

At higher orders we expect framing corrections to keep factorizing, generating exponentials. Therefore, we argue that the framing dependence of the single-node contribution is the same as the one present in the Wilson loop expectation value, see \eqref{eq:bosonic_framing}. Combining the two blocks, we can then write
\begin{equation}\label{eq:twoptbosonic}
\begin{split}
    &\llangle \cO(x_1) \cO(x_2) \rrangle_{1/6} \\
    & =\dot x_1^{\mu}\dot x_2^{\nu} \, \frac{e^{i\pi\left(\frac{ N_1}{k} + O(1/k^3)\right)\mathfrak{f}}\langle \cB(x_1)  \cB(x_2) \rangle_{\mathfrak{f}=0} +e^{-i\pi\left(\frac{N_2}{k} + O(1/k^3)\right)\mathfrak{f}}\langle \hat\cB(x_1) \hat\cB(x_2) \rangle_{\mathfrak{f}=0}}{e^{i\pi\left(\frac{ N_1}{k} + O(1/k^3)\right)\mathfrak{f}}\langle W_{\rm bos} \rangle_{\mathfrak{f}=0} + e^{-i\pi\left(\frac{N_2}{k} + O(1/k^3)\right)\mathfrak{f}}\langle\hat W_{\rm bos}\rangle_{\mathfrak{f}=0}},
    \end{split}
\end{equation}
where we have used \eqref{eq:bosonic_framing} to make the framing dependence explicit also in the normalization factor. 

Two comments are now in order. First of all, since the framing contributions come exclusively  from contractions inside the Wilson loop and never involve the operator insertion, the choice of $\cO$ is irrelevant for determining the framing phase. Therefore, the structure in \eqref{eq:twoptbosonic} should hold for any diagonal operator $\cO$ of the form \eqref{eq:OperatorA}.  Secondly, already at one loop the normalized two-point function is framing-dependent. In fact, even if the same phases arise in the numerator and denominator, the two-node structure of the bosonic Wilson loop prevents  their cancellation. This conclusion holds also in the  ABJM limit, $N_1=N_2 \equiv N$. %In this case, expanding numerator and denominator at one loop and recalling that $\langle A A\rangle = - \langle\hat A \hat A\rangle$, we obtain
%\begin{equation}
%    \llangle \cO(x_1) \cO(x_2) \rrangle_{1/6}\bigg|_{\rm ABJM} = \left(1 + \frac{2i\pi N}{k}\mathfrak{f}\right)\dot x_1^{\mu} \dot x_2^{\nu} \, \langle A_{\mu}(x_1)A_{\nu}(x_2)\rangle + O\left(1/k^2\right).
%\end{equation}

%%%%%%%%%%%%%%%%%%%%%%%%%%

\subsection{Fermionic defect}

Moving on to fermionic defects, we compute the two-point correlation function of the same operator $\cO$ in \eqref{eq:OperatorA} on a fermionic 1/6 BPS Wilson loop corresponding to $\beta^j= \bar\beta_j =0$ in \eqref{eq:1/24-superconnection}-\eqref{eq:spinors}. 

In this case we need to take into account additional one-loop fermion exchanges. Precisely, for each gauge diagram with a collapsible propagator mentioned above we have an analogous diagram with a fermion propagator between points $x_3$ and $x_4$ on the Wilson loop. Again, adding vanishing contributions we manage to factorize the operator insertion and reconstruct the full Gauss linking integral. Using results in \cite{Bianchi:2024sod} for fermionic diagrams at generic framing and exponentiating the one-loop result, we eventually obtain 
\begin{equation}\label{eq:1/6BPScorrfunction}
\begin{split}
    &\llangle \cO(x_1)\cO(x_2)\rrangle_{\alpha\bar\alpha}  \\
    &=\dot x_1^{\mu}\dot x_2^{\nu} \, \frac{e^{\frac{i\pi}{k}(N_1-N_2\alpha\bar\alpha)\mathfrak{f}}\langle  \cB(x_1) \cB(x_2) \rangle_{\mathfrak{f}=0}+e^{\frac{i\pi}{k}(N_1\alpha\bar\alpha-N_2)\mathfrak{f}}\langle \hat \cB(x_1)  \hat \cB(x_2) \rangle_{\mathfrak{f}=0}}{\langle W(\alpha\bar\alpha)\rangle} \, ,
    \end{split}
\end{equation}
where $\langle W(\alpha\bar\alpha)\rangle$ is the expression in \eqref{eq:exponentialization} with $\beta^j= \bar\beta_j =0$. The contributions proportional to $\alpha\bar\alpha$ in the framing phases are due to fermion exchanges. 

We note that, as long as $\alpha_i,\bar\alpha^i$ are arbitrary, the framing dependence does not cancel. However, setting $\alpha\bar\alpha=1$, thus landing on the $1/2$ BPS Wilson loop, the two phases in the numerator become equal and cancel against the same phase from the denominator
\begin{equation}\label{eq:halfBPScorrfunction}
\begin{split}
    \llangle \cO(x_1)\cO(x_2)\rrangle_{1/2} &= \dot x_1^{\mu}\dot x_2^{\nu}\, \frac{e^{\frac{i\pi(N_1-N_2)}{k}\mathfrak{f}} \left(\langle  \cB(x_1)  \cB(x_2) \rangle_{\mathfrak{f}=0}+\langle \hat \cB(x_1)  \hat \cB(x_2) \rangle_{\mathfrak{f}=0}\right)}{e^{\frac{i\pi(N_1-N_2)}{k}\mathfrak{f}}\langle W_{1/2}\rangle_{\mathfrak{f}=0}} \\\ &= 
    \llangle \cO(x_1)\cO(x_2)\rrangle_{1/2} \Big|_{\mathfrak{f}=0} \, .
\end{split}
\end{equation}
Thus, the correlation function is framing independent. Though this has been checked only at one loop, we expect this result to be valid at any order. Despite lacking a rigorous proof of this expectation, the structure of the correlation function at strong coupling \cite{Correa:2023thy} offers compelling evidence in favor of it. We come back to this point in section \ref{sec:strongcoupling}, where we discuss the interpretation of framing at strong coupling.

The same calculation can be performed for generic operators of the form 
\begin{equation}
    \cO = \begin{pmatrix}
        \cB & \bar{\mathcal{F}} \\ \mathcal{F} & \hat \cB
    \end{pmatrix}
\end{equation}
where $\cB, \hat \cB$ are bosonic and  $\bar{\mathcal{F}}, \mathcal{F}$ are fermionic. At one loop, the framing pattern is the same, and we can write in general
\begin{equation}\label{eq:2ptganeral}
\begin{split}
    \llangle \cO(x_1)\cO(x_2) \rrangle_{\alpha\bar\alpha} = \frac{1}{\langle W(\alpha\bar\alpha) \rangle}&\bigg[e^{\frac{i\pi}{k}(N_1-N_2\alpha\bar\alpha)\mathfrak{f}}\left(\langle \cB(x_1)  \cB(x_2) \rangle_{\mathfrak{f}=0} + \langle \bar{\mathcal{F}}(x_1) \mathcal{F}(x_2) \rangle_{\mathfrak{f}=0} \right) \\&+e^{\frac{i\pi}{k}(N_1\alpha\bar\alpha-N_2)\mathfrak{f}}\left(\langle \hat\cB(x_1)  \hat\cB(x_2) \rangle_{\mathfrak{f}=0} + \langle {\mathcal{F}}(x_1) \bar{\mathcal{F}}(x_2) \rangle_{\mathfrak{f}=0} \right)\bigg].
\end{split}
\end{equation}

In conclusion, the above analysis shows that correlation functions of local operator insertions on the defect theories defined by the Wilson loops considered in this paper are generally framing-dependent. 

A notable exception is the maximally supersymmetric 1/2 BPS Wilson loop, for which both weak and strong coupling calculations (to be further discussed below) consistently indicate the absence of framing dependence, provided the correlators are properly normalized.

Among these two-point functions, those of the displacement operator are especially significant. At the superconformal fixed points, their coefficients act as central charges, governing the response of defects to contour deformations. As such, they form part of the conformal data of the defect theories. Our analysis shows that for the framed circular 1/2 BPS defects, any framing dependence in these central charges is canceled by the normalization of defect correlators. In contrast, the perturbative corrections to the central charges of the circular 1/6 BPS defect acquire a framing dependence that cannot be eliminated by taking the modulus. For the supersymmetry preserving case of $\mathfrak{f}=1$, the resulting coefficients generally develop imaginary contributions, signaling a loss of unitarity in the defect theory. The emergence of such imaginary terms is similar to the findings of \cite{Closset:2012vg,Closset:2012vp} for the free energy of $\mathcal{N}=2$ Chern-Simons theories.

%%%%%%%%%%%%%%%%%%%%%%%%%%%%%

\subsection{One-point function of the defect stress tensor}

In this section we consider correlation functions of the defect stress tensor as diagnostics for anomalies of the underlying conformal and supersymmetry invariance. In one dimension the stress tensor trivially coincides with its trace, implying it is identically zero at the RG fixed points corresponding to the $W_{1/6}$ and $W_{1/2}$ defects. Instead, when the system is perturbed away from a fixed point by a marginally relevant operator $\hat{d}$, a non-trivial stress tensor $T_D = \beta \hat{d}$ arises, where $\beta$ is the beta-function of the deformation parameter.

In our specific case, marginally relevant deformations are encoded in supermatrix operators. Considering for instance the flow in figure \ref{subfig:flowbranch1} from the bosonic $W_{1/6}$ to the fermionic $W_{1/2}$ defects, the deformation operator is read from equations \eqref{eq:W}--\eqref{eq:spinors} by setting $\beta^j = \bar\beta_j=0$. 

To avoid cluttering the notation, without losing generality we simplify the following analysis by setting $\alpha_1=\bar\alpha^1=0$ and $\alpha_2\equiv\alpha$, $\bar\alpha^2=\bar\alpha$. The result can be straightforwardly generalized to arbitrary parameters.
Taking into account the non-trivial beta-functions for the deformation parameters, along the flow the stress tensor evaluates to
\begin{equation}
\label{eq:Td}
    T_D = - \begin{pmatrix}
        \beta_{\alpha\bar\alpha} \, \frac{4\pi i}{k}C_2\bar C^2 & \beta_{\alpha}\,\eta \bar\psi^1 \\ 
        \beta_{\bar \alpha}\,\xi \psi_1  & \beta_{\alpha\bar\alpha} \, \frac{4\pi i}{k}\bar C^2 C_2
        \end{pmatrix}
\end{equation}
where $\beta_\alpha, \beta_{\bar \alpha}$ can be read in \eqref{eq:beta} and $\beta_{\alpha \bar \alpha}=  \alpha \beta_{\bar \alpha} + \bar{\alpha} \beta_\alpha$.

We first argue that $T_D$ is cohomologically trivial. To this end, we recall that, by construction, the superconnection $\cal L$ of the deformed Wilson loop is obtained from the superconnection ${\cal L}_0$ of the $W_{1/6}$ superconformal fixed point as 
\begin{equation}
    \mathcal{L}= \mathcal{L}_0 + i [{Q},G] + G^2\,,
\end{equation}
where $Q$ is a mutually preserved supercharge whose explicit expression is not necessary here,\footnote{Using the supercharges in \cite{Castiglioni:2022yes}, it is given by ${Q}=(Q_{12}^+ - iS_{12}^+) + (Q_{34+}-iS_{34+})$.} and $G$ is the following supermatrix
\begin{equation}
    G = \sqrt{-\frac{4\pi i}{k}}\begin{pmatrix}
        0 & \bar\alpha\, C_2 \\
        \alpha\, \bar{C}^2 & 0
    \end{pmatrix}\,.
\end{equation}
Using the beta-functions in \eqref{eq:beta} with $\bar\beta_k\beta^k=0$,
the defect stress tensor \eqref{eq:Td} can be rewritten as
\begin{align}
\begin{split}
    T_D 
    = \frac{N_1+N_2}{2k}(\alpha\bar\alpha-1)\left[2(\mathcal{L}-\mathcal{L}_0) + i[{Q}, G]\right]\,.
\end{split}
\end{align}
Finally, introducing the covariant supercharge $\cQ \equiv Q - G$, such that $[{Q}, G]= [\cQ, G]$, and taking into account that 
$\mathcal{L}-\mathcal{L}_0$ is $\cQ$-exact \cite{Gorini:2022jws},\footnote{In \cite{Gorini:2022jws} this was proven for Wilson loops defined on the straight line. However, under a conformal mapping, the same holds for circular loops.} we can write 
\begin{equation}
\label{eq:Qexact}
T_D = [\cQ, \cO ] \, , 
\end{equation}
being $\cO$  a local supermatrix operator whose explicit expression is not relevant here. 

With these premises, we now  study correlation functions of $T_D$ on the defect. We start by computing the one-point expectation value $\llangle T_D\rrangle$. Exploiting the properties of the covariant supercharge $\cQ$ and its action on Wilson lines, we can write
\begin{align}\label{eq:onept}
        \llangle T_D\rrangle = \llangle [\mathcal{Q},\cO]\rrangle = \langle Q \Big( W(2\pi,\tau)\cO(\tau)W(\tau,0) \Big)\rangle\, .
\end{align}

In the absence of anomalies, this identity is simply the manifestation of the relation between scale and supersymmetry invariance, and it is consistent with superconformal invariance being preserved at the fixed points. In fact, conformal symmetry implies the vanishing of the l.h.s., while the r.h.s. is zero because of supersymmetry invariance, $Q\left|0\right\rangle=0$. 
Away from the fixed points, but along enriched flows which preserve the supercharge $Q$, equation \eqref{eq:onept} implies $\llangle T_D \rrangle = 0$, although $T_D \neq 0$. This can be used as an alternative definition of enriched RG flows. 

Given that supersymmetry is preserved at the quantum level for $\mathfrak{f}=1$, we now investigate how this feature is reflected in the integrated one-point function of $\llangle T_D\rrangle$, as defined in \eqref{eq:onept}. At one loop, this quantity is directly proportional to the one-loop correction to the defect expectation value, precisely
\begin{equation}
    \int_0^{2\pi} d\tau\llangle T_D(\tau) \rrangle^{(1\textrm{L})} = 2(\alpha\bar\alpha-1)\langle 
    W(\alpha\bar\alpha)\rangle^{(1\textrm{L})} = -\alpha\bar\alpha(\alpha\bar\alpha-1) \frac{4\pi^2}{k}N_1N_2 \, \epsilon\,,
\end{equation}
in terms of the bare parameters. Replacing them with the renormalized ones as in \eqref{eq:renormalization}, we eventually obtain a one-loop contribution which vanishes for $\epsilon \to 0$, plus a finite two-loop contribution
\begin{equation}
    -\alpha\bar\alpha(2\alpha\bar\alpha-1)(\alpha\bar\alpha-1)\frac{2\pi^2}{k^2}N_1N_2(N_1+N_2) \, , 
\end{equation}
which needs to be combined with genuine two-loop diagrams. 

At two loops, contributions come from three diagrams which are again proportional to the first three diagrams in \eqref{eq:matterdiagrams} contributing to the Wilson loop expectation value. Exploiting those results, we obtain schematically
\begin{equation}
\begin{split}
    {\rm double~fermion}&=4(\alpha\bar\alpha-1)\int d\tau_{1>2>3>4}\Tr\langle L_F(\tau_1) L_F(\tau_2) L_F(\tau_3) L_F(\tau_4)\rangle \\ &=\alpha^2\bar\alpha^2(\alpha\bar\alpha-1)\frac{2\pi^2}{k^2}N_1N_2(N_1+N_2)(3-\mathfrak{f}^2)\,,
\end{split}
\end{equation}
\begin{equation}
\begin{split}
    {\rm vertex}&=2(\alpha\bar\alpha-1)\int d\tau_{1>2>3}\Tr\langle L_F(\tau_1) L_F(\tau_2) L_0(\tau_3)\rangle + \text{permutations} \\ &= -4\alpha\bar\alpha(\alpha\bar\alpha-1)\frac{4\pi^2}{k^2}N_1N_2(N_1+N_2)\,,
\end{split}
\end{equation}
\begin{equation}
\begin{split}
    {\rm gauge-fermion}&= 2(\alpha\bar\alpha-1)\int d\tau_{1>2>3>4} \langle L_F(\tau_1)L_F(\tau_2)L_0(\tau_3)L_0(\tau_4) \rangle + \text{permutations} \\ &= \alpha\bar\alpha(\alpha\bar\alpha-1)\frac{2\pi^2}{k^2}N_1N_2(N_1+N_2)\mathfrak{f}^2,
\end{split}
\end{equation}
where $L_0$ and $L_F$ are the diagonal (bosonic) and off-diagonal (fermionic) parts of the superconnection \eqref{eq:1/24-superconnection}, respectively.
Combining all the contributions, we eventually obtain
\begin{equation}\label{eq:stresstensor1ptfunction}
    \int_0^{2\pi} d\tau \llangle T_D(\tau) \rrangle^{(2\textrm{L})} = \beta_\alpha \beta_{\bar\alpha}\frac{2\pi^2}{k}N_1N_2(N_1+N_2)(1-\mathfrak{f}^2) \, .
\end{equation}

Crucially, as long as the framing $\mathfrak{f}$ is different from one, the $T_D$ expectation value on the left hand side is different from zero, thus signaling breaking of supersymmetry, $Q|0\rangle \neq 0$ in \eqref{eq:onept}. Instead, supersymmetry is restored choosing $\mathfrak{f} = 1$. This is in perfect agreement with the cohomological equivalence of Wilson loops holding at framing one, as previously discussed. 

Proceeding further, we can extend the results above to infer the stress tensor $n$-point function $\llangle T_D(x_1) \ldots T_D(x_n)\rrangle$ for generic $n$. At framing one, this expression should vanish identically. In fact, thanks to the cohomology ${\cal Q}^2 =0$, any string of $T_D$'s is ${\cal Q}$-exact. Therefore, due to the identity \eqref{eq:onept}, its expectation value on the Wilson loop is zero. However, we expect a supersymmetry breaking in any $n$-point function when evaluated away from framing one. This allows to conclude that, in the perturbative regime, $n$-point stress-tensor correlation functions have to be proportional to powers of $(1-\mathfrak{f}^2)$, for any $n$.

%% file: 5Defect.tex
\section{The \emph{g}-theorem for dCFTs with framing}\label{sec:gtheorem}

A $g$-theorem for line defects in $d$-dimensional field theories has been proven in \cite{Cuomo:2021rkm}. It states that RG flows connecting different defect conformal field theories (dCFTs) are characterized by a monotonically decreasing function $g$, which coincides with the defect partition function at the fixed points, such that $g_\text{UV}>g_\text{IR}$. The theorem was formulated and proven for completely general defects. In this section, we study whether and how the presence of non-trivial framing affects its validity, a phenomenon specific to three-dimensional Chern–Simons theories and not incorporated in the original derivation. This analysis relies on our perturbative results at generic framing and it is consistent with general Ward identities in the presence of framing, as we are going to discuss.

To be specific, we focus on the RG flow connecting the bosonic $1/6$ BPS loop to the $1/2$ BPS one, obtained by setting $\beta^j = \bar\beta_j =0$ in \eqref{eq:W}–\eqref{eqn:M124bos}. As can be seen from \eqref{eq:renormalizedvev}, the framing-$\mathfrak{f}$ expectation value of this branch of operators at two loops is given by
\begin{equation}\label{eq:W1/6alpha}
    \begin{split}
    \langle W \rangle_\mathfrak{f} & = 1 +\frac{i\pi }{k}(N_1-N_2)\,\mathfrak{f} \\
    & \quad - \frac{\pi^2}{6k^2}\bigg[ 
    (1+3\mathfrak{f}^2)(N_1-N_2)^2-1 -3N_1N_2(\alpha\bar\alpha-1)^2(1-\mathfrak{f}^2) -2N_1N_2 \bigg]\, .
\end{split}
\end{equation}
Here, $\alpha\bar\alpha$ is the running effective parameter satisfying the one-loop beta-function equation 
\begin{equation}
\label{eq:compositebeta}
   \mu \frac{\partial (\bar\alpha\alpha)}{\partial\mu} = \alpha\bar\alpha(\alpha\bar\alpha-1)\frac{(N_1+N_2)}{k}\,,
\end{equation}
obtained from \eqref{eq:beta}.

A scheme-independent $g$-function decreasing along the RG flow was identified with  the defect entropy \cite{Cuomo:2021rkm}
\begin{equation}\label{eq:defectentropy1}
    g = \left( 1 +\beta_{\alpha\bar\alpha}\frac{\partial}{\partial (\alpha\bar\alpha)} \right)\log |\langle W \rangle|\, .
\end{equation}
At the fixed points it coincides with the defect partition functions $g_\text{UV} \equiv \log |\langle W_{1/6}\rangle|$ and $g_\text{IR} \equiv \log |\langle W_{1/2}\rangle|$, respectively. To prove the monotonicity of this function it is sufficient to establish the sign of $\mu \frac{d g}{d\mu}$, where $\mu$ is the renormalization scale.  

For a circular Wilson loop of radius $R$, the derivative of the defect entropy with respect to the energy scale can be written in terms of the defect stress tensor $T_D$, as follows \cite{Cuomo:2021rkm}
\begin{equation}\label{eq:1}
    \mu \frac{d g}{d\mu} = -R^2 \int d\tau_{1>2} \llangle T_D(\tau_1) T_D(\tau_2)\rrangle+ R \int d\tau_1 \llangle T_D(\tau_1)\rrangle\,.
\end{equation}
Using a crucial identity which relates the integrated one- and two-point functions 
\begin{equation}\label{eq:2}
    R\int d\tau_1 \llangle T_D(\tau_1)\rrangle = R^2 \int d\tau_{1>2} \llangle T_D(\tau_1) T_D(\tau_2)\rrangle \cos\tau_{12}\,,
\end{equation}
one obtains the scaling equation 
\begin{equation}\label{eq:g-theorem}
    \mu \frac{d g}{d\mu} = -R^2 \int d\tau_{1>2} \llangle T_D(\tau_1) T_D(\tau_2) \rrangle(1-\cos\tau_{12})\, .
\end{equation}
This allows to conclude that for unitary theories -- where $\llangle T_D(\tau_1) T_D(\tau_2) \rrangle >0$ -- the quantity $\mu \dfrac{d g}{d\mu}$ is always negative. Therefore,  $g$ is monotonically decreasing along the RG flow and $g_\text{UV} > g_\text{IR}$. 

Equation \eqref{eq:2} follows from the general Ward identity 
\begin{equation}
\label{eq:WI}
    \langle Q_\xi(\cD) \rangle= \int d^2\Sigma^\mu \langle T^b_{\mu\nu}\rangle \xi^\nu  = 0 \,,
\end{equation}
which establishes the invariance of the asymptotic vacuum under the action of conformal $SL(2, \mathbb{R})$ transformations generated by charges $Q_\xi$  defined on a bi-dimensional surface $\Sigma^\mu$ wrapping the defect $\cD$. Here $T^b_{\mu\nu} = -\tfrac{2}{\sqrt{g}} \tfrac{\delta S}{\delta g_{\mu\nu}}$ is the bulk stress tensor satisfying the  Ward identity for diffeomorphism invariance, $\nabla^\mu  T^b_{\mu\nu} = - \delta_D^{(2)} n_i^\nu D^i$, and $\xi^\nu$ are the conformal Killing vectors
\begin{equation}\label{eq:killing}
\begin{split}
    \xi_{(a)}^{\mu} &= \frac{1}{2}\left[ \delta^\mu_a (R +x^2/R)- 2x^\mu x_a / R \right],  \qquad a = 1,2 \, ,\\
    \xi_\phi^\mu & = \delta^\mu_a \epsilon^{ab}x_b\, .
\end{split}
\end{equation}
The first two generate linear combinations of translations and special conformal transformations on the defect plane, whereas the third one corresponds to rotations around the axis of the circle. 

In the presence of a relevant deformation which drives the defect out of the fixed point, the one-dimensional conformal group is broken down to the translations along the defect. However, this can be compensated by a suitable transformation of the dilaton $\Phi \to \Phi + \delta \Phi$, where $\Phi$  fixes the renormalization scale of the theory ($T_D = \tfrac{\delta W}{\delta \Phi}$). Therefore, the Ward identity \eqref{eq:WI}, with $T^b_{\mu\nu}$ now sourced also by the dilaton 
\begin{equation}
\label{eq:conservation}
    \nabla^\mu  T^b_{\mu\nu} = - \delta_D^{(2)} \left( n_i^\nu D^i + (\dot{T}_D - \dot{\Phi} T_D ) \right) \, , 
\end{equation}
can be interpreted as stating the equivalence between two different defects corresponding to two different dilaton fields. Expanding the identity $\log{|W_\Phi|} = \log{|W_{\Phi + \delta \Phi}|}$ in powers of $\delta \Phi$ around $\Phi=0$ eventually allows to obtain the crucial identity \eqref{eq:2} (see \cite{Cuomo:2021rkm, Shachar:2022fqk} for details). 

A natural question is whether the inclusion of non-trivial framing affects the validity of the $g$-theorem.
The first evidence comes from a two-loop evaluation of the one- and two-point functions in \eqref{eq:1}, at generic framing. 
The one-point function has been considered in the previous section and its two-loop expression is given in \eqref{eq:stresstensor1ptfunction}. For the two-point function, we can use the results in \cite{Castiglioni:2023uus} to obtain
\begin{equation}
     \int d\tau_{1>2} \llangle T_D(\tau_1) T_D(\tau_2) \rrangle\cos\tau_{12} = \beta_{\alpha}\beta_{\bar\alpha}\frac{2\pi^2}{k}N_1N_2(N_1+N_2)\,,
\end{equation}
where the $\beta$-functions on the right-hand side of this equation are given in \eqref{eq:beta}.
A direct comparison of the two results shows that, in the presence of framing, identity \eqref{eq:2} is modified as
\begin{equation}
\label{eq:2modified}
    R\int d\tau_1 \llangle T_D(\tau_1)\rrangle = R^2 \int d\tau_{1>2} \llangle T_D(\tau_1) T_D(\tau_2)\rrangle \cos\tau_{12} \, (1-\mathfrak{f}^2) \, .
\end{equation}
Consequently, at non-trivial framing the scaling equation \eqref{eq:g-theorem} becomes 
\begin{equation}\label{eq:g-theoremframing}
\begin{split}
    \mu \frac{d g}{d\mu} & = -R^2 \int d\tau_{1>2} \llangle T_D(\tau_1) T_D(\tau_2) \rrangle\left[1-\cos\tau_{12}(1-\mathfrak{f}^2)\right] \\
    &  \simeq \beta_\alpha\beta_{\bar\alpha} \frac{N_1+N_2}{2k}(1-\mathfrak{f}^2) +  \dots \,.
    \end{split}
\end{equation}
The net effect, already visible at two loops, is that the $\mathfrak{f}^2$ term spoils the definite sign of this expression, thus leading to a violation of the $g$-theorem for $\mathfrak{f}^2 >1$, unless one redefines $g$ as $g \sign{\!(1 - \mathfrak{f}^2)}$.

\begin{figure}
    \centering
    \includegraphics[width=0.7\linewidth]{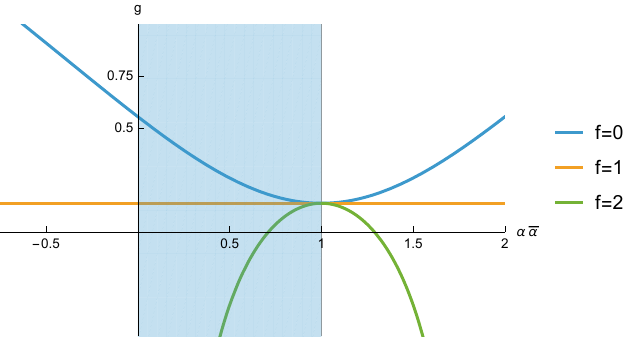}
    \caption{The $g$-function for the $1/6$ BPS operator depending on $\alpha\bar\alpha$, see \eqref{eq:W1/6alpha}, for different values of framing $\mathfrak{f}$. The blue region is bounded by the two fixed points: $\alpha\bar\alpha=0$ is a UV fixed point and $\alpha\bar\alpha=1$ in an IR fixed point.}
    \label{fig:g-function1}
\end{figure}
In figure \ref{fig:g-function1} we show the plot of $g$ as a function of $\alpha\bar\alpha$, for different values of framing. We highlight in blue the range of $0 \leq \alpha\bar\alpha \leq 1$, since the two limiting values correspond to the UV and IR fixed points of the RG flow for $\alpha\bar\alpha=0$ and $\alpha\bar\alpha=1$, respectively. For $\mathfrak{f}=1$ the $\alpha\bar\alpha$ dependence drops out and the $g$-function becomes constant, as expected from the cohomological equivalence (all defects along the RG flow have the same partition function $\log{|\langle W \rangle|}$). We stress that there is still a flow for $\mathfrak{f}=1$, since the beta functions are non-vanishing regardless of framing. This simply means that the VEV of the WL is not an appropriate observable to diagnose the flow. For $0\le \mathfrak{f}<1$,\footnote{In principle $\mathfrak{f}$ is an integer, but here we consider its analytic continuation to the real axis. Manifestation of non-integer framings already appeared in \cite{Bianchi:2014laa} for Wilson loops defined on latitudes.} the function $g$ is monotonically decreasing as a function of $\alpha\bar\alpha$, in agreement with the $g$-theorem, as increasing $\alpha\bar\alpha$ from 0 to 1 follows the RG flow from the UV to the IR fixed points. This means that $g$ is a monotonically increasing function of the energy scale, which is consistent with \eqref{eq:g-theoremframing}, observing that $\beta_\alpha \beta_{\bar\alpha}\geq0$ in the regime of interest.
Conversely, for $\mathfrak{f}>1$ $g$ is increasing as a function of the coupling $\alpha\bar\alpha$, signaling the aforementioned breakdown of the expected behavior from the $g$-theorem. At $\alpha\bar\alpha=1$ $g$ becomes framing-independent, because in the $1/2$ BPS case the framing contribution appears as an overall phase, which cancels out when taking the modulus of $\langle W \rangle$. 

While at the technical level the violation of the $g$-theorem can be clearly traced back to the non-trivial dependence of the expectation value $\langle W \rangle$ on $\mathfrak{f}$, it is interesting to examine how this breakdown arises directly within the proof of \cite{Cuomo:2021rkm}.

The form of identity \eqref{eq:WI} further manipulated using the Ward identity \eqref{eq:conservation} holds under the assumption that the conformal symmetry transformations leave the defect invariant, including its normal bundle -- a condition naturally satisfied when the normal bundle is trivial. However, a non-trivial normal bundle, such as the one defined by a necessarily non-planar framing contour $\Gamma_{\mathfrak{f}}$ winding $\mathfrak{f}$ times around the circle $\Gamma$, is not invariant under these transformations. 
 
In such a situation, the invariance of the defect expectation value $\langle \cD \rangle$ under an infinitesimal diffeomorphism generalizes as \cite{McAvity:1993ue, Billo:2016cpy} 
\begin{equation}\label{eq:Q}
    \delta \langle  \cD  \rangle = \int_{\cal M}  d^3 x  \, \langle T_{\mu\nu} \,  \cD \rangle \delta g^{\mu\nu} + \int_\cD d\tau \left(  \llangle \lambda^i_\mu\rrangle \delta n^\mu_i + \frac{1}{2} \llangle C_i \rrangle \delta K^i \right)+ \ldots  = 0 \, ,
\end{equation}
where $n^\mu_i$, $i=1,2$ are  two unit vectors normal to the defect, $K^i$ is the scalar curvature in the two normal directions, and $ T_{\mu\nu} = -\tfrac{\delta S}{\delta g^{\mu\nu}}$, $ \lambda^i_\mu = -\tfrac{\delta S}{\delta n_i^\mu}$, $C_i =   -\tfrac{\delta S}{\delta K_i}$, with $S$ being the total action of the bulk theory plus the defect.
The ellipsis denotes terms that give rise to the r.h.s. of \eqref{eq:conservation},  plus additional contributions that will not be relevant for the present discussion. 

For a framed defect, this leads to a generalization of the stress tensor conservation law \eqref{eq:conservation}, 
which now includes extra terms in the r.h.s. proportional to  defect correlation functions like $\llangle \lambda^i_\mu\rrangle$ and  $\llangle C_i \rrangle$. These new terms will necessarily leave an imprint in the proof of the $g$-theorem, in particular in the Ward identity \eqref{eq:WI}, thus leading to a non-trivial dependence on $\mathfrak{f}$ in the scaling equation for $g$. Here we provide a qualitative discussion of this effect. A rigorous proof would require evaluating perturbatively all terms in \eqref{eq:Q}, something which is beyond the scope of the present analysis. 

We limit our discussion to the study of the contribution arising from $\llangle C_i\rrangle$ in \eqref{eq:Q}. To evaluate it, we consider the framed contour $ \Gamma_\mathfrak{f}$ to be a helix of radius $\delta$ parametrized as
\begin{equation}\label{eq:helix}
   x^\mu(\tau) = (\cos\tau,\sin\tau,0) + \delta \, (\cos(\mathfrak{f}\tau)\cos\tau, \cos(\mathfrak{f}\tau)\sin\tau,\sin(\mathfrak{f}\tau))\,,
\end{equation}
such that framing effects can be recovered in the $\delta\to 0$ limit, as typically done in the literature \cite{Bianchi:2016yzj,Bianchi:2024sod}.

The geometry of the helix is described by the one-dimensional einbein $e^\mu = \dot{x}^\mu$. In flat space and for $\delta = 0$ the one-dimensional metric is trivial, in fact $\gamma = \dot{x}^\mu \dot{x}^\nu \delta_{\mu \nu} = 1$. We define the normal bundle to be given by the two unit vectors 
\begin{equation}
\begin{split}
& n_1^\mu = (\cos(\mathfrak{f}\tau)\cos\tau, \cos(\mathfrak{f}\tau)\sin\tau,\sin(\mathfrak{f}\tau)) \\
& n_2^\mu = (\sin(\mathfrak{f}\tau)\cos\tau, \sin(\mathfrak{f}\tau)\sin\tau,-\cos(\mathfrak{f}\tau)) \, , 
\end{split}
\end{equation}
satisfying $n_i^\mu n_{j \mu} = \delta_{ij}, n_i^\mu \dot{x}_\mu =0$. 
In this parametrization the extrinsic curvatures along the two directions normal to the helix are defined by the equation $\partial_\tau e^\mu = n_i^\mu K_i$ \cite{Billo:2016cpy}, which eventually leads to  $K_1=-\cos\mathfrak{f}\tau, K_2 = -\sin\mathfrak{f}\tau$ at leading order in $\delta$. Under the action of an infinitesimal transformation generated by Killing vectors \eqref{eq:killing} 
they transform as 
\begin{equation}
\label{eq:variations}
            \delta_{\xi_{(1)}} K_i  =  -\sin\tau \partial_\tau K_i  \, ,  \qquad \delta_{\xi_{(2)}} K_i =  \cos\tau \partial_\tau K_i \, , \qquad
            \delta_{\xi_\phi} K_i = -\partial_\tau K_i 
\end{equation}

At $\mathfrak{f}=0$ the scalar curvature becomes constant, $K_1= -1, K_2 = 0$ and its variation vanishes. Instead, in the general case $K_i$ depends explicitly on the contour parameter $\tau$ and a non-vanishing contribution potentially appears in the right hand-side of \eqref{eq:Q}, depending on $\llangle C_i \rrangle$ being vanishing or not. 

In order to evaluate $\llangle C_i \rrangle$, we first decompose the vector field $A_\mu$ along the tangent and normal directions to the defect, 
\begin{equation}
    A_\mu = \dot x_\mu (\dot x^\nu A_\nu) + n_\mu^i (n_i^\nu A_\nu) \equiv \dot x_\mu A_\tau + n_\mu^i A_i
\end{equation}
and similarly for the derivatives $\partial_\mu$. Inserting into the ABJ(M) action, after some algebra the Chern-Simons term reduces to
\begin{eqnarray}\label{eq:CSaction}
    S_{A\partial A} &=& \frac{k}{4\pi} \int d^3x \epsilon_{\mu\nu\rho}\bigg[ \dot x^\mu A_\tau A_\lambda \left( n_1^\nu n_2^\rho n_1^\lambda - n_2^\nu n_1^\rho n_2^\lambda \right) \mathfrak{f}  \cr
    && - \dot x^\mu n_i^\nu n_j^\rho \Big( A_\tau n_j^\sigma \partial_i A_\sigma + A_iA_\tau K_j - A_i n_j^\sigma \partial_\tau A_\sigma - A_j \dot x^\sigma \partial_i A_\sigma \Big) \bigg]. 
\end{eqnarray}
This expression exhibits a non-trivial dependence on the scalar curvature $K_j$. Taking into account that the defect superconnection does not depend on $K_i$, we  eventually  obtain 
\begin{equation}\label{eq:Ci}
    C_i =  -  \frac{\partial S}{\partial K_i}   =  -  \frac{\partial S_{A\partial A}}{\partial K_i}   = \frac{k}{2\pi}\epsilon_{\mu\nu\rho} \dot x^\mu n_j^\nu n_i^\rho A^j A_\tau \, .
\end{equation}

We then evaluate $\int d\tau \llangle C_i \rrangle \delta_\xi K^i$ in \eqref{eq:Q}, by inserting the above expression for $C_i$, the $\delta_\xi K^i$  variations given in \eqref{eq:variations} and using definition \eqref{eq:correlators} for the one-point correlation function on the defect. 

At one loop, contributing diagrams come from contracting gauge fields in $C_i$, which sit at point $\tau$, and gauge fields from the expansion of the Wilson loop, which sit at points $\tau_1$ and $\tau_2$. The integrand is explicitly given by
\begin{equation}
    I(\tau,\tau_1,\tau_2)\equiv \frac{k}{2\pi}\epsilon_{\mu\nu\rho}\dot x^\mu n_j^\nu n_i^\rho  \langle (n_j^{\alpha}A_\alpha(\tau))(\dot x^{\beta}A_\beta(\tau)) (\dot x_1^{\lambda} A_{\lambda}(\tau_1))(\dot x_2^{\gamma} A_{\gamma}(\tau_2)) \rangle \,\delta_{\xi} K^i \,.
\end{equation}
Once we take into account the two possible contractions and insert the corresponding gauge propagators, $I(\tau,\tau_1,\tau_2)$ becomes
\begin{equation}
    \frac{k}{2\pi}\epsilon_{\mu\nu\rho} \dot x^{\mu}  n_j^\nu n_i^\rho n_j^\alpha \dot x^\beta  \epsilon_{\alpha\lambda \sigma} \epsilon_{\beta\gamma\eta} \bigg( \frac{\dot x_1^\lambda(x-x_1)^\sigma}{|x-x_1|^3} \frac{\dot x_2^\gamma(x-x_2)^\eta}{|x-x_2|^3} +  \frac{\dot x_1^\gamma(x-x_1)^\eta}{|x-x_1|^3} \frac{\dot x_2^\lambda(x-x_2)^\sigma}{|x-x_2|^3}  \bigg) \delta_\xi K^i \,.
\end{equation}

\begin{figure}[H]
    \centering
    \includegraphics[width=0.7\linewidth]{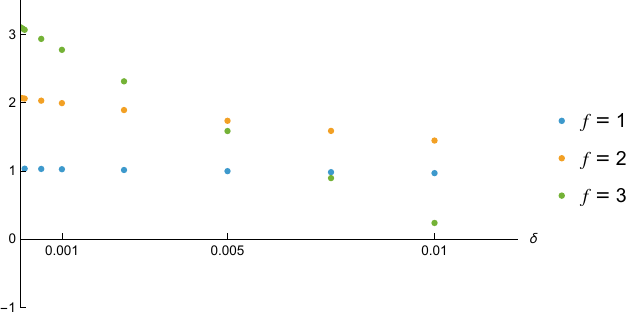}
    \caption{One-loop contribution of the integrated one-point function of $C_i$ as a function of $\delta$, for different values of framing $\mathfrak{f}$. }
    \label{fig:integratedC}
\end{figure}

Now, considering that gauge fields coming from the Wilson loop expansion can appear in three different places -- both fields before $C_i$, one before and one after,  or both after $C_i$ -- the full contribution is the sum of the three different integrals
\begin{align}\label{eq:integratedC}
    \begin{split}
        \int_0^{2\pi} d\tau \left( \int_0^{\tau} d\tau_1 \int_0^{\tau_1} d\tau_2   + \int_\tau^{2\pi} d\tau_1 \int_\tau^{\tau_1} d\tau_2   + \int_\tau^{2\pi} d\tau_1 \int_0^{\tau} d\tau_2 \right)  I(\tau,\tau_1,\tau_2)\,.
    \end{split}
\end{align}
The exact evaluation of this integral is a hard task. However, since we are only interested in understanding whether the extra terms in \eqref{eq:Q} can be responsible for $\mathfrak{f}$-dependent contributions to the $g$ scaling equation, we limit to providing numerical evidence that \eqref{eq:integratedC} is framing-dependent. To do so, we regularize potential short distance divergences in \eqref{eq:integratedC} by taking $\tau$ to parametrize the ordinary circle, 
\begin{equation}\label{eq:circle}
    x^\mu(\tau) = (\cos\tau,\sin\tau,0)\,,
\end{equation}
while $\tau_1$ and $\tau_2$ parametrize two distinct toroidal helices  \cite{Bianchi:2024sod}
\begin{equation}
    x^\mu_{k=1,2} = (\cos\tau_k,\sin\tau_k,0) + k\,\delta \, (\cos(\mathfrak{f}\tau_k)\cos\tau_k, \cos(\mathfrak{f}\tau_k)\sin\tau_k,\sin(\mathfrak{f}\tau_k))\,.
\end{equation}
These toroidal helices have distinct infinitesimal radius ($\delta$ and $2\delta$) but both wind around the circle in \eqref{eq:circle} $\mathfrak{f}$ times. Using these contour parametrizations, we evaluate the integral in \eqref{eq:integratedC} numerically as a function of the infinitesimal radius $\delta$, and for different values of the framing $\mathfrak{f}$. Choosing for instance the transformations under the $\xi_\phi$ Killing vector, the results are plotted in figure \ref{fig:integratedC}, where it appears that in the $\delta\to0$ limit a remnant framing-dependent contribution survives.
Though this is not a complete proof, it gives a clear indication that for framed defects we should expect the scaling equation for $g$  in \eqref{eq:g-theorem} to be affected by extra $\mathfrak{f}$-dependent terms originating from the generalized Ward identity \eqref{eq:Q}. This is consistent with our perturbative findings. 

\vskip 10pt

We close this section with a brief comment about the second branch of RG flows, that is the  branch described by the set of parameters $\{\beta^j,\bar\beta_j\}$ ($j=3,4$) with $\alpha_i = \bar\alpha^i=0$ (see figure \ref{subfig:flowbranch2}). From \eqref{eq:renormalizedvev} the corresponding expectation value is given by
\begin{equation}
\label{eq:W1/6beta}
\begin{split}
    \langle W \rangle_{\mathfrak{f}} = 1 + \frac{i\pi}{k}(N_1-N_2)\,\mathfrak{f} - \frac{\pi^2}{6k^2}\bigg[& (1+3\mathfrak{f}^2)(N_1-N_2)^2-1 \\&-3N_1N_2\left( \bar\beta\beta+1\right)^2 (1-\mathfrak{f}^2) - 2N_1N_2 \bigg].
\end{split}
\end{equation}

The two-point function of the defect stress tensor has been evaluated in \cite{Castiglioni:2023uus}. The result shows that operators within this branch represent defect theories that are not reflection-positive (in Euclidean space) or, equivalently, unitary (in Minkowski space). 

\begin{figure}[H]
    \centering
    \includegraphics[width=0.7\linewidth]{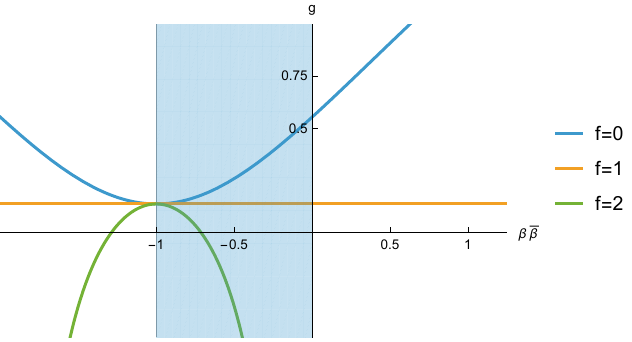}
    \caption{The $g$-function for the $1/6$ BPS operator depending on $\beta\bar\beta$, see \eqref{eq:W1/6beta}, for different values of framing $\mathfrak{f}$.
    The blue region is bounded by the two fixed points: $\beta\bar\beta=-1$ is a UV fixed point and  $\beta\bar\beta=0$ in an IR fixed point.}
    \label{fig:g-function2}
\end{figure}

For defect theories that do not possess reflection positivity, the $g$-theorem is not expected to hold. Nevertheless, as done for the first branch of operators, it is interesting to see how the $g$-function behaves for different values of framing. In figure \ref{fig:g-function2} we plot $g$ for different values of $\mathfrak{f}$. The blue region is bounded by UV and IR fixed points ($\beta\bar\beta=-1$ and $\beta\bar\beta=0$, respectively). For $\mathfrak{f}=0$, we see that $g_{\rm UV}<g_{\rm IR}$, in accordance with the framing zero studies \cite{Castiglioni:2023uus}. On the other hand, the correct monotonicity of $g$ seems to be recovered for $\mathfrak{f}>1$ (green line). This peculiar connection between framing and unitarity is quite interesting and would require further investigation. 

%% file: 6strongcoupling.tex
\section{Framing at strong coupling}
\label{sec:strongcoupling}

The strong coupling description of Wilson loops in ABJ(M) via holography has been widely discussed in the literature, see for example  \cite{Drukker:2008zx,Chen:2008bp,Rey:2008bh,Berenstein:2008dc, Drukker:2009hy} and chapters 12-13 of \cite{Drukker:2019bev} for a review. However, to the best of our knowledge an investigation of framing in this context is still missing. In this section we advance a proposal for the holographic dual of framing, which we claim to be given by the coupling of the string dual to the Wilson loop to the background Kalb-Ramond field of the theory.

ABJM theory \cite{Aharony:2008ug} is dual to M-theory on $AdS_4 \times S^7/\mathbb{Z}_k$, which at large $k$ reduces to type IIA string theory on $AdS_4\times \mathbb{CP}^3$
\begin{equation}
     ds^2 = R^2\left( ds^2_{AdS_4} + 4 ds^2_{\mathbb{CP}^3} \right)\,,
\end{equation}
where $R^2/\alpha' =\pi\sqrt{2\left(\lambda-\frac{1}{24}\right)}$ is the radius of the space as a function of the 't Hooft coupling $\lambda = N/k$, including an anomalous shift \cite{Bergman:2009zh}, which is subleading and negligible in the present calculation.
We take the $\mathbb{CP}^3$ to be described by the Fubini-Study metric
\begin{equation}
\label{metricCP3}
\begin{split}
    ds^2_{\mathbb{CP}^3} =& \frac{1}{4}\bigg[d\alpha^2 + \cos^2\tfrac{\alpha}{2}\left( d\theta_1^2 + \sin^2\theta_1 d\varphi^2_1 \right)+\sin^2\tfrac{\alpha}{2}\left( d\theta_2^2 + \sin^2\theta_2 d\varphi^2_2 \right) \\ &  \hskip .5cm + \sin^2\frac{\alpha}{2}\cos^2\frac{\alpha}{2}\left( d\chi +\frac{\cos\theta_1}{2} d\varphi_1 -\frac{\cos\theta_2}{2} d\varphi_2  \right)^2\bigg]\,,
\end{split}
\end{equation}
where $0\le\alpha,\theta_1,\theta_2\le\pi$, $0\le \varphi_1,\varphi_2 \le 2\pi$ and $0\le\chi\le 2\pi$. In the ABJ case, the background also includes a Kalb-Ramond field \cite{Aharony:2008gk,Bergman:2009zh,Aharony:2009fc}
\begin{equation}\label{eq:B-field}
    B^{(2)} = \frac12\left(\frac{N_1-N_2}{k}+\frac12\right) dA,
\end{equation}
with the K\"ahler potential $A$ of $\mathbb{CP}^3$ given by
\begin{equation}
\label{Kaehler potential}
    A = \cos\alpha d\chi + 2\cos^2\tfrac{\alpha}{2}\cos\theta_1 d\varphi_1 + 2\sin^2\tfrac{\alpha}{2}\cos\theta_2 d\varphi_2.
\end{equation}
The flux of this field over $\mathbb{CP}^1 \subset \mathbb{CP}^3$ gives the difference in ranks, $|N_1-N_2|$, of the gauge fields at the two nodes of the quiver.

To identify the holographic dual of framing it is useful to consider two specific cases: the 1/2 BPS loop discussed repeatedly above and the 1/6 BPS latitude  \cite{Cardinali:2012ru,Bianchi:2014laa,Bianchi:2018bke,Griguolo:2021rke} (see chapters 7-8 of \cite{Drukker:2019bev} for a review). 

The $1/2$ BPS loop in the fundamental representation is dual to a fundamental string spanning an $AdS_2$ subspace of $AdS_4$ and sitting at a point in $\mathbb{CP}^3$ \cite{Drukker:2009hy}, so to preserve an $SU(3)$ subgroup of the R-symmetry. This configuration can be obtained by setting $\alpha=0$ and $\theta_1 = 0$ in \eqref{metricCP3}. On the other hand, the 1/6 BPS latitude is supported on a latitude of $S^2$ and has a string theory dual \cite{Correa:2014aga} constructed by setting $\varphi_1=\tau$ (with $\tau$ being the parameter along the loop, identified in static gauge with one of the world-sheet coordinates) and allowing $\theta_1$ to vary over the other worldsheet coordinate $\sigma$, with its boundary value fixed to be the latitude angle, $\theta_1(\sigma=0)=\theta_0$. The $1/2$ BPS loop solution is then retrieved in the limit in which the latitude goes to the equator of the sphere ($\theta_0\to 0$ in our parametrization). 

Thinking of the 1/2 BPS loop as a special case of the latitude, we set $\varphi_1=\tau$ also in this case. This does not change the solution itself, since the pull-back of the $\varphi_1$-direction contains a factor of $\sin\theta_1$, which is vanishing for the 1/2 BPS loop. The string is still localized at a point in the internal space.\footnote{In particular, we consider $\theta_1$ as the polar angle and $\varphi_1$ as the azimuthal angle in the $\mathbb{CP}^1$ subspace of $\mathbb{CP}^3$ parameterized by $(\theta_1,\varphi_1)$. Then $\theta_1=0$ identifies the North pole of $S^2\simeq \mathbb{CP}^1$ where the value of $\varphi_1$ is arbitrary.} However, even if the string solution is unchanged, if $d\varphi_1$ does not vanish, the pullback of the K\"ahler potential \eqref{Kaehler potential} onto the string world-sheet becomes non-trivial. In particular, for the latitude Wilson loop (which also has $\alpha=0$ and $\chi$ constant), it reads
\begin{equation}\label{eq:KahlerPotLat}
    A = 2\cos\theta_1d\varphi_1 = 2 \cos\theta_1 d\tau\,,
\end{equation}
reducing to $A=2d\tau$ in the $1/2$ BPS limit. 

This gives rise to a non-trivial Kalb-Ramond field \eqref{eq:B-field}, which then couples to the string solution, contributing to its classical action. The minimal surface contribution, for which only the $AdS$ part of the spacetime is relevant in the 1/2 BPS case, evaluates as usual to $\pi\sqrt{\lambda_1+\lambda_2}$, with $\lambda_{1,2}=N_{1,2}/k$. 
The coupling to the $B^{(2)}$ field generates, on top of this, an imaginary term from the Wess-Zumino piece of the Euclidean action
\begin{equation}
    S_{\textrm{B}} = \frac{i}{2} \int_{AdS_2} B^{(2)}.
\end{equation}
Using \eqref{eq:B-field} and Stokes' theorem to evaluate the integral we obtain
\begin{equation}
    S_\textrm{B} = \frac{i}{2} \left( \frac{N_1-N_2}{k} + \frac{1}{2} \right) \int_{\partial AdS_2} d\tau \cos\theta_1\Big|_{\partial AdS_2} = i \pi \left( \frac{N_1-N_2}{k} + \frac{1}{2} \right )\cos\theta_1\Big|_{\partial AdS_2}.
\end{equation}
Including this contribution to the minimal surface term above, we find the expectation value
\begin{equation}
   \left\langle W_{1/2} \right\rangle \simeq e^{i\pi\frac{N_1-N_2}{k}}e^{\pi\sqrt{\lambda_1+\lambda_2}}\,.
\end{equation}
The phase factor coincides with the framing phase at $\mathfrak{f}=1$ obtained at weak coupling via a perturbative computation, and at any coupling via localization \cite{Drukker:2010nc}. 

As a non-trivial check of this identification, we note that it correctly reproduces the framing phase also for the latitude. In fact, in this case the minimal surface term together with the B-field contribution give
\begin{equation}
   \left\langle W_{\mathrm{latitude}} \right\rangle\simeq e^{i\pi\frac{N_1-N_2}{k}\nu}e^{\pi\sqrt{\lambda_1+\lambda_2}\nu},
\end{equation}
where $\nu\equiv \cos\theta_1\Big|_{\partial AdS_2}=\cos\theta_0$. This matches precisely the perturbative result of \cite{Bianchi:2014laa} and the exact result obtained from the latitude matrix model  \cite{Bianchi:2018bke}.

It is instructive to explicitly realize that the holographic computation of the Wilson loops selects automatically the correct framing choice, $\mathfrak{f}=1$ for the maximal circle and $\mathfrak{f}=\nu$ for the latitude, as required by supersymmetry \cite{Kapustin:2009kz,Bianchi:2014laa}.\footnote{From the discussion above, modifying the internal rotation by $\varphi_1=k\,\tau$ would generate a phase proportional to $k$, mimicking a framing-$k$ factor at weak coupling. However, the classical area also changes non-trivially, so the configuration no longer corresponds to the weak-coupling operators under consideration and would typically break supersymmetry.} This is, of course, not surprising if one recalls our previous discussion of how framing is needed to cancel cohomological anomalies at the quantum level: we confirm here, in the deep quantum regime of the gauge theory operators probed by holography, what was also signaled at the first orders in perturbation theory by the computations of section \ref{sec:perturbative}.

The bosonic $1/6$ BPS Wilson loop has been argued to be dual to a string solution smeared over a $\mathbb{CP}^1$ equator of $\mathbb{CP}^3$, which suitably reduces the amount of preserved supersymmetry. Since its position in $\mathbb{CP}^3$ is not fixed, this has been interpreted as the string solution obeying Neumann rather than Dirichlet boundary conditions in the internal space. Solutions with mixed boundary conditions interpolate between the $1/2$ and $1/6$ BPS dual configurations \cite{Correa:2019rdk,Garay:2022szq}.

From the matrix model calculation \cite{Drukker:2010nc}, averaging over the $\mathbb{CP}^1$ effectively multiplies the $1/6$ BPS expectation value at strong coupling by a factor proportional to the volume of $\mathbb{CP}^1$. Such prefactors, which multiply the dominant exponential behavior of Wilson loops at strong coupling, are typically difficult to extract from a direct string calculation. Nonetheless, the observed effect aligns with the smearing interpretation. Importantly, this averaging should not modify the exponential part of the expectation value, which arises from evaluating the classical string action.
We therefore expect that the same reasoning applies to the part of the action coupling to the $B$-field, which should govern the framing dependence at strong coupling. This leads to a prediction for the framing phase of the $1/6$ BPS Wilson loop at strong coupling, which is $e^{i\pi(\lambda_1 - \lambda_2)}$.

We finally test our identification of framing at strong coupling against a localization calculation. Following the steps of \cite{Marino:2009jd,Drukker:2010nc}, while keeping $N_1 \neq N_2$, we find 
\begin{align}\label{eq:1/2bpsStrongCoupling}
    \langle W_{1/2} \rangle_{\mathfrak{f}=1} &= \frac{e^{i \pi B}}{8\pi \hat\lambda} \kappa(\hat\lambda,B)\,, \cr
\langle W_{\text{bos}} \rangle_{\mathfrak{f}=1} &\simeq -\frac{e^{i\pi B}\kappa}{4\pi^2 i \lambda_1}\left(\log\kappa-1-i\pi B\right),
\end{align}
where we emphasize that the calculation is performed at framing 1.  
In the above formulae, $B = \frac{N_1 - N_2}{k} + \frac{1}{2}$, $\hat\lambda = \frac{N_1 + N_2}{2k}$, and the overall normalization is chosen to match unit tree level expectation values at weak coupling. The expression for the 1/2 BPS Wilson loop is exact, whereas the expectation value of the 1/6 BPS Wilson loop has already been restricted to large $\kappa$.
The real function $\kappa$ must be expanded in the relevant regime, in order to retrieve expressions depending on the 't Hooft parameters. At strong coupling, it asymptotes to $\kappa \sim e^{\pi\sqrt{2\hat\lambda}}$, reproducing the classical string area.

In conclusion, we highlight the role of $B$ in the result, which in the string theory picture corresponds to the flux of the Kalb--Ramond two-form. It fully characterizes the effect of framing at strong coupling, triggering the emergence of imaginary contributions. For the $1/2$ BPS Wilson loop it is a simple phase which coincides with the weak coupling prediction. Conversely, in the $1/6$ BPS case the same phase appears with additional corrections.
On the one hand, this localization result validates our string-theoretic interpretation of framing. On the other hand, it resonates with weak coupling findings for the $1/6$ BPS Wilson loop, hinting at a non-trivial framing dependence of its expectation value \cite{Bianchi:2016yzj}.

As a further confirmation of our identification of the holographic dual of framing, we note that  it also agrees with the framing dependence of defect correlation functions, computed in section \ref{sec:corrfunctions}. In fact, in \cite{Correa:2023thy} it was shown that in the case of Dirichlet boundary conditions in the internal space, \textit{i.e.} for the dual of the $1/2$ BPS Wilson loop, correlation functions at strong coupling do not depend on $B$. On the other hand, in the case of Neumann boundary conditions, \textit{i.e.} for the bosonic $1/6$ BPS Wilson loop, correlation functions do exhibit a non-trivial dependence on $B$.
These strong coupling findings are perfectly consistent with our weak coupling analysis. In fact, we have found that  framing drops out from the $1/2$ BPS Wilson loop, see \eqref{eq:halfBPScorrfunction}, while it does not cancel for the $1/6$ BPS case in \eqref{eq:1/6BPScorrfunction}.

%% file: Conclusions.tex
\section{Conclusions}
\label{sec:conclusions}

In this paper we have been concerned with framing effects in the evaluation of BPS Wilson loops in ABJ(M) theory, both at weak and at strong coupling. 

At weak coupling our main result is the generalization of our perturbative two-loop computation \cite{Bianchi:2024sod} of the 1/2 BPS Wilson loop at generic framing to the parametric 1/24 BPS operator. We provide a direct check of the cohomological equivalence at framing one of all the operators interpolated by this loop. The expectation value \eqref{eq:renormalizedvev} depends generically on the eight parameters $\{\alpha_i,\bar\alpha^i,\bar\beta_j,\beta^j\}$, but this dependence drops entirely when $\mathfrak{f}=1$, as expected from localization. 

In the direction of strengthening checks between perturbative and localization results, a natural generalization of this analysis is to consider parametric $\theta_0$-latitudes \cite{Castiglioni:2022yes}. Also in that case, one expects the parametric dependence to disappear and agreement with localization to be found for $\mathfrak{f}=\nu=\cos\theta_0$, instead of $\mathfrak{f}=1$. This was confirmed at one loop in \cite{Bianchi:2024sod}, but a higher-loop computation is still missing. 

A second result in perturbation theory has been the evaluation of correlation functions of local operators inserted on framed Wilson loops, highlighting the framing-dependent contributions to these quantities, see \eqref{eq:twoptbosonic} for a bosonic defect and \eqref{eq:2ptganeral} for a fermionic one. Interestingly, these contributions drop in the 1/2 BPS case. Moreover, we have considered the integrated one-point function of the defect stress tensor \eqref{eq:stresstensor1ptfunction}, which turns out to be proportional to $(1-\mathfrak{f}^2)$, at least up to two loops in perturbation theory. This offers an alternative perspective on the fact that $\mathfrak{f}=1$ preserves supersymmetry and the cohomological equivalence at the quantum level. 

Finally, we have discussed how framing affects defect Ward identities and the $g$-theorem proved in \cite{Cuomo:2021rkm}. We have found that the relation \eqref{eq:2} between the integrated one- and two-point functions of the defect stress tensor, which is crucial in the proof of the $g$-theorem, has to be modified as in \eqref{eq:2modified}. The inclusion of the $(1-\mathfrak{f}^2)$ framing dependence spoils the $g$-theorem for $\mathfrak{f}^2>1$. This modification can be tracked back to new terms appearing in the defect Ward identity, which are due to the fact that, in the presence of framing, the normal bundle to the defect is no longer trivial. This is encoded in \eqref{eq:Q}. We have focused on just one term in that expression, showing that a framed defect does indeed introduce framing dependence in the Ward identity.

Our analysis completes the general picture of defect RG flows in ABJ(M) initiated in \cite{Castiglioni:2022yes}. According to the definition of anomaly given in \cite{Closset:2012vg,Closset:2012vp}, a theory is anomalous whenever the whole set of its classical properties cannot be required simultaneously at the quantum level. In our case the properties that appear to be incompatible are the cohomological equivalence of the spectrum of BPS defects under investigation and framing independence, that is independence of the trivialisation of the frame bundle. Therefore, we have three options:
\begin{itemize}
    \item We give up cohomological equivalence and avoid introducing a framing dependence. This is the approach used in \cite{Castiglioni:2022yes,Castiglioni:2023uus,Castiglioni:2023tci} where we studied RG flows at framing zero (ordinary perturbation theory). 
    \item We give up framing independence setting the theory at framing one, thus restoring cohomological equivalence. This is the canonical approach of localization. As already discussed, BPS defects still flow but their expectation value is no longer an appropriate observable to detect the flow. 
    \item We give up both cohomological equivalence and framing independence. This is what we have done in the RG flow analysis in this paper. This less conservative option allows to better investigate the role of framing in the definition of defects, thus leading to a deeper comprehension of the special role played by framing one. 
\end{itemize}

In the end, the legitimate question one should ask is whether framing is physical or not. Our present understanding is that framing should be seen as part of the definition of quantum defect observables, even though the bulk theory in which the defects are embedded is insensible to it, and rightly so. In other words, from the point of view of the ABJ(M) theory, framing is nothing but a regularization prescription, but for defects it discriminates between different quantum behavior. Therefore, defects at different framing should be considered as different one-dimensional theories. 
However, to have a stronger confirmation of this interpretation, the current spectrum of results should be completed with information from the holographic description of defects. 

At strong coupling, we have proposed a holographic dual to framing, namely the coupling of the fundamental string to the background B-field of the ABJ theory. We have checked that this yields the correct framing phase both in the case of the 1/2 BPS circle and the 1/6 BPS latitude, naturally selecting the values $\mathfrak{f}=1$ and $\mathfrak{f} = \nu$, respectively. Whether Wilson loops allow for a consistent dual description also at $\mathfrak{f}\neq 1$ or $\mathfrak{f}\neq \nu$, we do not yet know. If our interpretation of framing is correct, the answer should be “yes”, as Wilson loops at different framing should correspond to physically distinct observables with some broken symmetries, like supersymmetry. This is still an open question, which urgently asks for further investigation. 

An interesting extension of this research would be to consider Wilson loops in higher representations of the gauge group, both on the gauge theory side using perturbation theory and on the string theory side. In the latter case, the dual objects would be D-branes, rather than fundamental strings, which could provide a further check of our proposal in section \ref{sec:strongcoupling}.

%%%%%%%%%%%%%%%%%

\subsection*{Acknowledgments}

We are grateful to D. Correa, N. Drukker, A. Faraggi, Y. Papadimitriou and G. Silva for comments and discussions. DT thanks the organizers and participants of the Favignana {\it Workshop on higher-dimensional integrability} for a stimulating meeting, where parts of this paper were presented. This work was supported in part by the INFN grant {\it Gauge and String Theory (GAST)}. DT would also like to thank FAPESP’s partial support through the grant 2019/21281-4. MB is supported by Fondo Nacional de Desarrollo Cient\'ifico y Tecnol\'ogico, through Fondecyt Regular 1220240, Fondecyt Exploraci\'on 13220060 and Fondecyt Exploraci\'on 13250014. MT is supported by the Simons Foundation through the award number 1023171-RC.